\documentclass[12pt]{book}
\usepackage[pdftex]{graphicx}
\usepackage{amssymb}
\usepackage{amsmath}
\usepackage[]{enumerate}
\usepackage{braket}
\usepackage{bm}

\newcommand{\me}{\mathrm{e}}

\newcommand{\mi}{\mathrm{i}}
\newcommand{\dif}{\mathrm{d}}
\newcommand{\al}{\alpha}
\newcommand{\ep}{\epsilon}

\newcommand{\p}{\partial}

\newcommand{\bro}{\bm{\rho}}

\DeclareBoldMathCommand{\bA}{A}
\DeclareBoldMathCommand{\bB}{B}
\DeclareBoldMathCommand{\bC}{C}
\DeclareBoldMathCommand{\bD}{D}
\DeclareBoldMathCommand{\bE}{E}
\DeclareBoldMathCommand{\bF}{F}
\DeclareBoldMathCommand{\bG}{G}
\DeclareBoldMathCommand{\bH}{H}
\DeclareBoldMathCommand{\bI}{I}
\DeclareBoldMathCommand{\bJ}{J}
\DeclareBoldMathCommand{\bK}{K}
\DeclareBoldMathCommand{\bL}{L}
\DeclareBoldMathCommand{\bM}{M}
\DeclareBoldMathCommand{\bN}{N}
\DeclareBoldMathCommand{\bO}{O}
\DeclareBoldMathCommand{\bP}{P}
\DeclareBoldMathCommand{\bQ}{Q}
\DeclareBoldMathCommand{\bR}{R}
\DeclareBoldMathCommand{\bS}{S}
\DeclareBoldMathCommand{\bT}{T}
\DeclareBoldMathCommand{\bU}{U}
\DeclareBoldMathCommand{\bV}{V}
\DeclareBoldMathCommand{\bW}{W}
\DeclareBoldMathCommand{\bX}{X}
\DeclareBoldMathCommand{\bY}{Y}
\DeclareBoldMathCommand{\bZ}{Z}
\DeclareBoldMathCommand{\ba}{a}
\DeclareBoldMathCommand{\bb}{b}
\DeclareBoldMathCommand{\bc}{c}
\DeclareBoldMathCommand{\bd}{d}
\DeclareBoldMathCommand{\be}{e}
\DeclareBoldMathCommand{\beff}{f}
\DeclareBoldMathCommand{\bg}{g}
\DeclareBoldMathCommand{\bh}{h}
\DeclareBoldMathCommand{\bi}{i}
\DeclareBoldMathCommand{\bj}{j}
\DeclareBoldMathCommand{\bk}{k}
\DeclareBoldMathCommand{\bl}{\ell}
\DeclareBoldMathCommand{\bem}{m}
\DeclareBoldMathCommand{\bn}{n}
\DeclareBoldMathCommand{\bo}{o}
\DeclareBoldMathCommand{\bp}{p}
\DeclareBoldMathCommand{\bq}{q}
\DeclareBoldMathCommand{\br}{r}
\DeclareBoldMathCommand{\bs}{s}
\DeclareBoldMathCommand{\bt}{t}
\DeclareBoldMathCommand{\bu}{u}
\DeclareBoldMathCommand{\bv}{v}
\DeclareBoldMathCommand{\bw}{w}
\DeclareBoldMathCommand{\bx}{x}
\DeclareBoldMathCommand{\by}{y}
\DeclareBoldMathCommand{\bz}{z}

\makeindex

\title{Primer on quantum weirdness}  
\author{R. D. Hazeltine}   

\begin{document}
\maketitle 
\tableofcontents
\mainmatter

\setcounter{chapter}{0}
\chapter{Introduction}\label{Ch0INTRO}
\section{Varieties of weirdness}
Quantum physics is weird for two distinct reasons. \begin{enumerate} \item The microworld differs profoundly from the macroworld of ordinary experience.  This demonstrable fact has nothing to do with quantum mechanics or any physical theory.  \item Quantum mechanics, the most successful description of the microworld, presents difficult theoretical issues.  It's self-consistency is not obvious.  \end{enumerate}  So the world is strange and the theory is strange.

\subsection{Microworld is strange} 
Prominent examples of phenomena peculiar to the atomic scale are briefly reviewed here. Most of these are discussed more thoroughly in later chapters.
\paragraph{Broken probability rules}
Suppose a certain event, $C$, cannot occur unless preceded by one of two mutually exclusive events, $A$ and $B$.  Then the probability of the occurrence of $C$ must be the sum of the probabilities for $A$ and $B$:
\[
P(C) = P(A) + P(B)
\]
The simplest situation contradicting this law is probably the double slit experiment, discussed in detail by Feynman\cite{feynman1965}. Experiments reveal  that ``interference'' terms must be included on the right-hand side.  We do not repeat the Feynman discussion here, but emphasize that the phenomenon is not an artifact of quantum mechanics or any other theory:  it is an experimental fact.

The usual probability rules apply in the macro-world because typical macroscopic systems have so many degrees of freedom---and so many closely spaced energy levels---to be  extremely sensitive to external perturbations.  Because the perturbations are essentially random, any realistic experimental resolution will average away the interference terms. One says that \emph{decoherence} eliminates interference. Of course special experiments, such as the double-slit, can be contrived to avoid decoherence.             

\paragraph{Non-additivity of experimental results}
Certain particles (spin-1/2) made to traverse an inhomogeneous magnetic field are observed to be deflected ``upwards'' (experimental value $+1$) or ``downwards'' (experimental value $-1$) with respect to the field direction. (The device producing this deflection was first constructed by Stern and Gerlach, and the experiment is named ``Stern-Gerlach.'' Various micro-phenomena are most simply illustrated by Stern-Gerlach experiments.) The deflection is always in one of these two directions, never in between, regardless of the orientation of the field.  The two resu lts $\pm 1$ are found to be equally likely, while the choice appears to be entirely random. So two immediate surprises appear: the randomness, perhaps the least troubling feature of the micro-world, and what is sometimes called ``spatial quantization,'' the appearance, for any orientation of the experimental device, of two discrete directions of deflection.  But probably the most mysterious feature of this class of experiments is the departure from additivity, discussed next.

For brevity we use quantum mechanical notation, without assu\-ming know\-ledge of the theory.  The experiment is said to measure ``spin,'' with components $(\sigma_{x}, \sigma_{y}, \sigma_{z})$. Each component is measured to be $\pm1$.  The sum of such spins, such as the quantity $\sigma_{k}  \equiv 2^{-1/2}(\sigma_{x} + \sigma_{y})$, must also be measurable, by a device oriented along a rotated axis.  Spatial quantization tells us that the only possible measured values for $\sigma_{k}$ are $\pm 1$. But its definition implies the allowed outcomes $\sigma_{k} = (2^{1/2}, 0, -2^{1/2})$.  

The quantum mechanical interpretation is straightforward, although the strangeness remains.  The point is that the eigenvalues of the sum of non-commuting hermitian operators, like $\sigma_{x}$ and $\sigma_{y}$, will differ from the sum of their eigenvalues.  Expectation values do not display this difference:  the sum of the expectation values is the expectation value of the sum.  As first pointed out by Grete Hermann \cite{hermann1935}, the von Neumann proof\cite{vonneumann1955} of the impossibility of non-dispersive ensembles errs in requiring additivity in any hidden-variable theory.

\paragraph{Failure of realism}
We have called the randomness of microscopic out\-comes---usually referred do as the \emph{indeterminism} of the micro-world---a least troubling feature of that world.  But one aspect of indeterminism requires emphasis.  One might assume that, before entering the Stern-Gerlach device, the particle had an unknown but definite orientation; the result is indeterminate only because certain ``hidden'' information is not available.  Remarkably, it can be shown that this assumption leads to false predictions: the particle carries no such information.  The direction of deflection is not predetermined, but realized only at the moment of deflection. 

An analogous situation is realized in the double-slit experiment.  There one can show that ascribing a definite trajectory to the particle (that is, which slit was used) leads to predictions contrary to experiment.\cite{feynman1965}

Such micro-phenomena suggest that measurements create reality: small things assume ``real'' properties only when observed.  An extreme version of this attitude is that the actual situation is undetermined until it is registered by a conscious mind.  However one can as well assume that certain macroscopic constructs (such as a Stern-Gerlach device or a sensitive screen) impose definite outcomes, without the need for an observing agent.

\paragraph{Failure of locality}
Consider two measuring devices, at two distant locations $A$ and $B$. Measurements at $A$ may be correlated with those at $B$, as occurs, for example, when both measurements involve particles that interacted in the past.  Such correlation would be unsurprising in the macro-world.  What is strange about micro-phenomena is that the observations at location $B$ may depend upon the state of the measuring equipment---upon which quantities have been selected for measurement---at location $A$. 

Because of quantum uncertainty, this non-local dependence cannot be used to send a message between $A$ and $B$: the changes at $B$ are statistical, and visible only after many repetitions. But the statistical influence is unmistakable and easily seen by \emph{post facto} examination of the full measurement record. Furthermore the influence of $A$'s equipment is transmitted instantaneously; an increasingly elaborate series of experiments sees no limitation to sub-luminal speeds. The conflict with relativity theory is muted because the non-local effect cannot be used for communication.  Nonetheless the fact of such an influence challenges the space-time locality assumed by relativity.

\subsection{The problem with quantum theory}
\paragraph{Collapse}
Quantum mechanics, mainly through its superposition principle, fits most of the micro-phenomena described above into a simple and compelling system.  It may not offer a vivid causal picture, \emph{a la} Newtonian mechanics, of microscopic events, but that failure is perhaps to be expected. What is offered is a powerful recipe for experimental predication, along with a fundamental illumination of such scientific domains as statistical mechanics and chemistry.  Yet despite these triumphs, despite the deep insights of Bohr, Heisenberg, Dirac, Schr\"{o}dinger and their contemporaries, quantum theory has a major flaw.

The essential problem is that quantum theory offers two distinct dynamical laws, without a clear criterion for choosing between them \cite{gisin1989, penrose1996, weinberg2015}.  One dynamic, the Schr\"odinger equation, is unitary and Hamiltonian; it is said to describe an isolated system.  The second dynamic is neither Hamiltonian nor unitary; it is said, at least in the early and most conventional treatments, to describe a system on which a ``measurement'' has been preformed \cite{vonneumann1955}.  A precise and general rule for determining what physical circumstance qualifies as a measurement remains vague and controversial (see \cite{BellBook} and references cited therein).  

It is not surprising that the intervention associated with measurement changes a system's state. What is significant here is, first, that the change is non-unitary, and outside the realm of Schr\"odinger evolution; it requires its own special dynamical law. Second, and perhaps more seriously, standard quantum mechanics lacks any clear, definitive statement as to when the non-unitary dynamic is called for.  For example, when a particle exiting a Stern-Gerlach device is observed to be deflected in a certain direction, a non-unitary evolution has certainly occurred.  But what if the deflection is not observed---not ``measured''?

The non-unitary evolution law has many labels: ``collapse of the wave function,'' ``reduction of the state vector,'' the quantum ``jump,'' and so on.  It is sometimes implied that the system changes from a superposition of states in Hilbert space to a single such state, but this idea does not make quantum-mechanical sense: all states are superpositions.  Indeed each observation has its own ``preferred basis'': the wave function collapses to a single state among the set of basis states preferred by the measuring device.

Various interpretations of quantum mechanics might make the theory more palatable, but they have no effect on its dynamical flaw.  For example  the collapse can be viewed as an essentially psychological event:  the state changes because new information has changed an observer's knowledge.  Yet the need for a well-defined dynamic is unchanged; any collapse has measurable effects that cannot be ignored. This point is emphasized by Bell and Nauenberg \cite{Bell1966a}, who state: ``Thus observation...is a dynamical interference with the system which may alter the statistics of subsequent measurements''. 

Quantum mechanics succeeds as a predictive tool because of its flawless description of equilibrium states and because,  in practice, its dynamical ambiguity is rarely an issue.  One usually knows whether and when a measurement has occurred, requiring the collapse dynamic.  There are exceptions to this practical rule, such as quantum evolution in the early universe, but its enormous triumphs---from statistical mechanics and chemistry to quantum field theory and quantum information theory---discourage challenge.

\paragraph{Perspectives}
The micro-world is so distant from the world in which brains evolved that it may never be understood, in any familiar sense of the word ``understand.''  Thus one should not attempt any picture of the micro-world, being instead content with a set of rules for predicting the probabilities of various experimental outcomes.  In this perspective wave function collapse is seen, not as a physical process, but as part of a recipe. 

Alternative viewpoints ascribe to collapse, and even to the wave function, a certain physical reality. Examples of this perspective include the pilot-wave theory \cite{debroglie, bohm1952}, the gravitational explanation for collapse\cite{penrose1996}, and the many-worlds interpretation\cite{everett1957}.  A striking example begins with the premise that traditional quantum dynamics is simply wrong. Its two conflicting dynamical laws are replaced by a single equation for wave-function evolution, with universal applicability.\cite{GRW1986}\\cite{GRW1990}cite{Adler2004}  The Lindblad equation\cite{linblad1976} often plays a role in such developments.

\subsection{The density operator}\label{density}
The density operator---its construction, nature and evolution---will be a central issue of this primer. This operator, originally called the density matrix \cite{vonneumann1955}, displays in its simplest (``proper'') form the state of a system as a linear combination of projection operators.  The coefficient of each projector is the probability for observation of the corresponding subspace. 

The density operator obviates certain subjective or anthropocentric aspects of quantum mechanics  \emph{if}\begin{enumerate}\item The goal of quantum theory is strictly taken to be the prediction of densities, rather than states.  It predicts only probabilities, not events. \item The density coefficients are unambiguously specified by the nature of some measuring device---by a well defined state preparation procedure. \end{enumerate} In this case quantum theory would not need to mention observers; measurement produces a certain density operator, not a certain observation.  The domain boundary of quantum mechanics is statistical prediction. 

A possible advantage of this point of view, related to the so-called statistical interpretation of quantum mechanics\cite{Ballentine1970}, is to reduce any psychological component in the quantum theory of measurement. Statistical predictions are in some sense more objective than observations. But the statistical interpretation does not in it self address the problem of collapse.  According to conventional quantum theory  the density operator is subject to two distinct laws of evolution: in its own way, it collapses. Theories that avoid the dual evolution law are considered in Chapter \ref{Ch6LIN}.


\section{Purpose of this work}
Treatises on our topic often propose a new interpretation, or even a new version, of quantum mechanics, intended to alleviate the weirdness.  We have a much smaller ambition: to introduce the key ideas and mathematical tools central to modern discussions of collapse. We assume knowledge of basic quantum mechanics---the Schr\"{o}dinger evolution of states in Hilbert space, the Born probability rule, the Dirac formalism, and so on---while explicating such ideas as projectors, density operators, Bell inequalities, entanglement and the Lindblad equation. In other words this work is a \emph{primer} in the simplest sense of that word: without taking a stand on the major questions, it is intended to make an enormous  and growing body of research more accessible.  

\setcounter{chapter}{1}
\chapter{Review of states and operators}\label{Ch1PRE}

\section{Basics}
\subsection{Dyads} \label{sec:label}
A dyad is an operator constructed from two state vectors; it is the outer product of a ket and a bra:
\[
D = \ket{\psi}\bra{\phi}
\]
It acts on kets according to
\[
D \ket{\chi} = \ket{\psi}\bra{\phi}\ket{\chi} 
\]
Using a familiar rule (adjoint of product is reversed product of adjoints) we find the Hermitian adjoint (hereafter, adjoint) operator
\[
D^{\dagger} = \ket{\phi}\bra{\psi}
\]

In a useful special case, we consider a complete orthonormal (CO) set of states $\ket{\chi_{\mu}}$, introduce the double index $\bj = (\mu, \nu)$, and define
\[
D_{\bj}  = \ket{\chi_{\mu}}\bra{\chi_{\nu}} 
\]
The double-index notation is convenient and helps to avoid certain confusions.  The adjoint operator is of course
\[
D_{\bj}^{\dagger} = \ket{\chi_{\nu}}\bra{\chi_{\mu}}
\]
The components of $D_{\bj}$, in the basis $\ket{\chi_{\mu}}$, are 
\[
(D_{\bj})_{\alpha \beta}  = \bra{\chi_{\al}}(\ket{\chi_{\mu}}\bra{\chi_{\nu}} )\ket{\chi_{\beta}} = \delta_{\al \mu}\delta_{\beta \nu}
\]

Finally we consider a linear combination
\[
D \equiv \sum_{\bj} d_{\bj}D_{\bj} = \sum_{\mu, \nu} d_{\mu \nu}  \ket{\chi_{\mu}}\bra{\chi_{\nu}} 
\]
This operator has components
\[
D_{ij} =  \sum_{\mu, \nu} d_{\mu \nu} \bra{\chi_{i}}D_{\mu, \nu}\ket{\chi_{j}} =\sum_{\mu, \nu} d_{\mu \nu} \delta_{i \mu} \delta_{j \nu} = d_{ij}
\]
It follows that $D$ is Hermitian (self-adjoint) iff the matrix $d_{ij}$ is Hermitian. 

It is somewhat instructive to show this fact more explicitly.  The adjoint of the operator $D$ is
\[
D^{\dagger} = \sum_{\bj} d^{*}_{\bj}D^{\dagger}_{\bj} = \sum_{\mu, \nu} d^{*}_{\mu \nu} \ket{\chi_{\nu}}\bra{\chi_{\mu}}
\]
or, after exchanging labels $\mu \leftrightarrow \nu$,
\[
D^{\dagger}  = \sum_{\mu, \nu} d^{*}_{\nu \mu} \ket{\chi_{\mu}}\bra{\chi_{\nu}}
\]
This sum agrees with that for $D$ provided $d^{*}_{\nu \mu}  = d_{\mu \nu}$, as stated.

\subsection{Projectors: definitions}\label{pdef}
The diagonal version $D_{\phi \phi}$ is a projection operator, or projector, which we denote by
\[
\mathcal{P}_{\phi} = \ket{\phi}\bra{\phi}
\]
It is clear that $\mathcal{P}_{\phi}$ is Hermitian, and that it satisfies the identity 
\[
\mathcal{P}_{\phi}^{2} = \mathcal{P}_{\phi}
\]
which defines a projector. It follows that the eigenvalues of a projector are either zero or unity.  

These properties deserve emphasis: a projector $\mathcal{P}_{\phi}$ is an Hermitian operator with the spectrum $(0,1)$. The eigenvalue $1$ corresponds to eigenstate $\phi$, while the (highly degenerate) eigenvalue $0$ has as its eigenstates all the other (orthogonal ) states belonging to $\mathcal{P}_{\phi}$.  

Our interest focuses on dyadic projectors, which project onto a single state.  More generally one can project onto a subspace $S$ of dimension $s>1$.  Most of the same algebra applies, but the trace is in general the dimension $s$ of the target space.  An example is the projector
\[
\mathcal{R}_{\phi} \equiv \bI - \mathcal{P}_{\phi}
\]
where $\bI$ is the identity.  The trace of this non-dyadic projector may be infinite.

When acting on any ket $\ket{\psi}$, the projector yields the ket 
\[
\mathcal{P}_{\phi} \ket{\psi} = \braket{\phi|\psi} \ket{\phi}
\]
It follows in particular that
\[
\braket{\psi | \mathcal{P}_{\phi} | \psi} = | \braket{\phi|\psi} |^{2}
\]
so that the dyadic projector is a positive operator, but not a unitary operator.

The product of two (dyadic) projectors vanishes iff they project onto orthogonal states.  Thus let $P_{i} = \ket{\chi_{i}}\bra{\chi_{i}}$ and consider
\[
P_{1}P_{2} \ket{\psi} = \ket{\chi_{1}}\braket{\chi_{1}|\chi_{2}}\braket{\chi_{2}|\psi}
\]
This will vanish for all $\psi$ iff $\braket{\chi_{1}|\chi_{2}} = 0$.  If the $\chi_{i}$ are not orthogonal then the projectors do not commute,
\[
[P_{1},P_{2}] \neq 0
\]
Indeed in the non-orthogonal case $P_{1}$ and $P_{2}$ produce different kets.

A linear combination of projectors is not in general a projector.  Thus the operator
\[
\mathcal{Q} \equiv  a\mathcal{P}_{\phi} + b\mathcal{P}_{\psi}
\]
where $a$ and $b$ are constants, is a projector only if \begin{enumerate} \item the states $\phi$ and $\psi$ are orthogonal, so that
\[
\mathcal{P}_{\phi}\mathcal{P}_{\psi} = 0 = \mathcal{P}_{\psi}\mathcal{P}_{\phi} 
\]
and \item $a^{2}  = a,\,\, b^{2} = b$. That is, both $a$ and $b$ must be zero or unity.\end{enumerate}  In other words we have, for orthogonal projectors $\mathcal{P}_{\mu}$,
\begin{equation}
\label{ppd}
\mathcal{P}_{\mu}\mathcal{P}_{\nu} = \delta_{\mu \nu}\mathcal{P}_{\mu}
\end{equation}
and the only linear combination that yields a projector has the form
\[
\mathcal{P}  = \mathcal{P}_{1} + \mathcal{P}_{2} + \cdots
\]
where the $\mathcal{P}_{i}$ are mutually orthogonal.  (If the sum includes a term for each of the CO set $\chi_{n}$, then this projector is the unit operator.)  

We often consider the set of  projectors onto a CO set $\chi_{i}$, using an abbreviated notation
\[
\mathcal{P}_{\chi_{i}} = \mathcal{P}_{i}
\]
The completeness of the set is expressed by
\begin{equation}
\label{sp1}
\sum_{i} \mathcal{P}_{i} = \bI
\end{equation}
where $\bI$ is the unit operator.

Any Hermitian operator $A$ can be expressed in terms of its projectors.  Hermiticity implies that $A$ has a CO set of eigenkets $\ket{\chi_{\nu}}$; denoting the eigenvalues by $a_{\nu}$ we have
\begin{equation}
\label{ev}
A \ket{\chi_{\nu}} = a_{\nu}\ket{\chi_{\nu}}
\end{equation}
which implies
\begin{equation}
\label{ap}
A = \sum_{\nu}a_{\nu}\ket{\chi_{\nu}}\bra{\chi_{\nu}} = \sum_{\nu}a_{\nu}\mathcal{P}_{\nu}
\end{equation}
The representation (\ref{ap}) is unique in the case of a non-degenerate spectrum.  In the case of a degenerate eigenvalue, one can as usual choose different combinations of eigenstates within each subspace.

To see that (\ref{ev}) implies (\ref{ap}), post-multiply both sides of the former by $\bra{\chi_{\nu}}$ to get $A\mathcal{P}_{\nu} = a_{\nu} \mathcal{P}_{\nu}$; then sum over $\nu$ to obtain
\[
A\sum_{\nu}\mathcal{P}_{\nu}= \sum_{\nu}a_{\nu}\mathcal{P}_{\nu}
\]
But the left-hand side of this relation is simply the operator $A$, so (\ref{ap}) follows.

An important extension is the statement that
\begin{equation}
\label{fa}
f(A) = \sum_{\nu}f(a_{\nu})\mathcal{P}_{\nu}
\end{equation}
for any smooth function $f$. Consider for example the function $f(A)  = A^{2}$:
\[
A^{2} = \sum_{\mu \nu}a_{\mu}a_{\nu}\mathcal{P}_{\mu}\mathcal{P}_{\nu}
\]
Now (\ref{ppd}) implies 
\[
A^{2} = \sum_{\mu \nu}a_{\mu}a_{\nu}\delta_{\mu \nu}\mathcal{P}_{\mu} = \sum_{\mu}a_{\mu}^{2}\mathcal{P}_{\mu} 
\]
as given by (\ref{fa}).  A similar manipulation verifies (\ref{fa}) when $f$ is any power, and therefore any polynomial.  Thus it can be assumed to hold for any sufficiently smooth function.

This is a convenient place to mention an identity similar to (\ref{fa}), but not involving projectors.  Suppose $A$ is a normal matrix with eigenvalues $a_{\nu}$, and denote by $A_{\mu\nu}$ the components of $A$ with respect to that basis in which $A$ is diagonal:
\[
A_{\mu\nu} = \delta_{\mu\nu}a_{\nu}
\]
(Remember that all sums are explicit in this discussion; we do not use the Einstein summation convention.). Then, in that special basis, the components of a function $f(A)$ are given by
\begin{equation}
\label{fan}
[f(A)]_{\mu\nu} = \delta_{\mu\nu}f(a_{\nu})
\end{equation}
The proof is completely analogous to that of (\ref{fa}): one considers the components of powers of $A$.

\subsection{Projector algebra}\label{proalg}
We consider the projector onto some state $\phi$,
\[
\mathcal{P}_{\phi} \equiv \ket{\phi}\bra{\phi}
\]
The trace of $\mathcal{P}_{\phi}$ is computed from a CO set $\chi_{n}$ as
\[
\text{Tr}(\mathcal{P}_{\phi}) = \sum_{n}\braket{\chi_{n}|\mathcal{P}_{\phi}|\chi_{n}} = \sum_{n} | \braket{\phi |\chi_{n}}|^{2} = \braket{\phi | \phi}
\]
Thus if $\phi$ is normalized we have
\[
 \sum_{n} | \braket{\phi |\chi_{n}}|^{2} = 1
\]
and 
\begin{equation}
\label{trp1}
\text{Tr}(\mathcal{P}_{\phi}) = 1
\end{equation}

\paragraph{Properties of trace} The trace is a linear map,
\[
\text{Tr}(\al \bA + \beta \bB) = \al \text{Tr}(\bA) + \beta \text{Tr}(\bB)
\]
that is invariant under transposition,
\[
\text{Tr}(\bA) = \text{Tr}(\bA^{T})
\]
under commutation,
\[
\text{Tr}(\bA\bB) = \text{Tr}(\bB  \bA)
\]
and (therefore) under similarity transformation
\[
\text{Tr}(\bS^{-1 } \bA \bS) = \text{Tr}(\bA)
\]
Invariance under commutation implies invariance under cyclic rearrangements of the triple product:
\[
\text{Tr}(\bA \bB \bC) = \text{Tr}( \bC \bB \bA) =\text{Tr}(\bB \bC \bA) 
\]
The trace of a Hermitian operator is real, since it has real eigenvalues, and invariant under Hermitian conjugation.  Importantly, the trace of a product of three Hermitian operators is invariant under arbitrary (not just cyclic) rearrangement.  The point is that if $\bA$, $\bB$ and $\bC$ are all Hermitian, then cyclic order can be broken:
\[
(\bA \bB \bC)^{\dagger}   = (\bB \bC)^{\dagger}\bA^{\dagger} = \bC \bB \bA
\]

\paragraph{Projector components}
In terms of the CO set $\ket{\chi_{n}}$, the components of any ket $\ket{\psi}$ are
\[
\psi_{m} = \braket{\chi_{m} | \psi}
\]
so that 
\[ 
\ket{\psi} = \sum_{n} \psi_{n} \ket{\chi_{n}}
\]
The state $\ket{\phi} = \bA \ket{\psi}$ has the components
\[
\phi_{m} = \bra{\chi_{m}}\bA \sum_{n}\psi_{n}\ket{\chi}_{n}  = \sum_{n}\psi_{n} \braket{ \chi_{m} |\bA | \chi_{n}}
\]
or 
\[
\phi_{m} = \sum_{n} A_{mn} \psi_{n}
\]
where
\[
A_{mn} = \braket{m | \bA | n}
\]

Now suppose that the operator is the projector onto some state $\psi$, $\bA = \ket{\psi} \bra{\psi}$.  We see that its components are
\begin{equation}
\label{proc}
\mathcal{P}_{\psi, mn} = \braket{\chi_{m} | \mathcal{P}_{\psi}  | \chi_{n}}  = \braket{\chi_{m} | \psi} \braket{\psi | \chi_{n}}  = \psi_{m}\psi_{n}^{*}
\end{equation}
A significant special case is that in which the basis states form a continuum: $\chi_{n} \rightarrow \chi(x)$. Then the projector components are
\[
\mathcal{P}_{\psi, xx'} = \psi(x)\psi^{*}(x')
\]
As a check, we compute
\[
(\mathcal{P}_{\psi}\ket{\phi})_{x} = \int dx' \mathcal(P)_{\psi, xx'}\phi(x')= \int dx'\, \psi(x)\psi^{*}(x')\phi(x')= \braket{\psi | \phi} \psi(x)
\]
That is 
\[
\mathcal{P}_{\psi}\ket{\psi} = \braket{\psi | \phi} \ket{\psi}
\] as before.

\subsection{Projector evolution}
To find the dynamics of the projector $\mathcal{P} \equiv \ket{\phi}\bra{\phi}$, we use the identity
\[
\mathcal{P} \ket{\psi} = \braket{\phi|\psi}\ket{\phi}
\]
to compute the operator $\p_{t}\mathcal{P}$. The Liebniz rule provides
\[
(\p_{t}\mathcal{P}) \ket{\psi} = (\p_{t}\ket{\phi})\braket{\phi |\psi} + \ket{\phi} (\p_{t}\bra{\phi}) \ket{\psi}
\]
The Schr\"odinger equation (with $\hbar = 1$)
\[
\p_{t}\ket{\phi} = - \mi H \ket{\phi}
\]
where $H$ is the system Hamiltonian, then implies
\begin{eqnarray*}
(\p_{t}\mathcal{P}) \ket{\psi} &=& - \mi H \ket{\phi}\braket{\phi |\psi} + \mi \ket{\phi} \braket{H\phi |\psi} \\
&=& -\mi H \mathcal{P} \ket{\psi} + \mi \ket{\phi} \braket{\phi| H |\psi} \\
&=& -\mi \left[ H \mathcal{P} - \braket{\phi| H |\psi}\right] \ket{\phi}
\end{eqnarray*}
Here the second term can be expressed as
\[
\braket{\phi| H |\psi} \ket{\phi} = \mathcal{P} H \ket{\psi}
\]
so we have
\[
(\p_{t}\mathcal{P}) \ket{\psi} = - \mi[H, \mathcal{P}] \ket{\psi}
\]
That is,
\begin{equation}
\label{cp}
\p_{t} \mathcal{P} = -\mi [H, \mathcal{P}]
\end{equation}
whence
\begin{equation}
\label{pt2}
\mathcal{P}(t) = \me^{-\mi Ht}\mathcal{P}(0) \me^{\mi H t}
\end{equation}

\newpage
\section{Spin algebra}\label{spinal}
\subsection{Spin 1/2}
In this two-dimensional space it is customary to use basis states $\ket{z+}$ and $\ket{z-}$.  A common notation is $\ket{z+} = \ket{1}$, $\ket{z-} = \ket{0}$.  In order to define and use Pauli spin matrices, we need to express our states in terms of column vectors and row vectors.  Thus  we write
\begin{equation}
\ket{1} =
\left(
\begin{array}{c}
  1    \\
  0 \\
\end{array}\right), \,\,\,
\ket{0} =
\left(
\begin{array}{c}
  0    \\
  1 \\
\end{array}\right) 
\end{equation}
The inner product is
\[
(a, \,\, b) \left(
\begin{array}{c}
  c    \\
  d \\
\end{array}\right) = ac + bd
\]
and the outer product is
\[
\left(
\begin{array}{c}
  a    \\
  b \\
\end{array}\right)(c, \,\,\,d) = \left(
\begin{array}{cc}
 ac & ad   \\
 bc & bd \\
\end{array}\right)
\]

Next we recall that
\begin{equation}
\label{xyz}
\ket{x\pm} = 2^{-1/2}(\ket{1} \pm \ket{0}), \,\, \ket{y\pm} = 2^{-1/2}(\ket{1} \pm \mi \ket{0})
\end{equation}
in order to write
\begin{eqnarray}
\ket{x+} &=& 2^{-1/2}
\left(
\begin{array}{c}
  1    \\
  1 \\
\end{array}
\right), \,\,\, \ket{x-} = 2^{-1/2}
\left(
\begin{array}{c}
  1    \\
  -1 \\
\end{array}
\right) \\
\ket{y+} &=& 2^{-1/2}
\left(
\begin{array}{c}
  1    \\
  \mi \\
\end{array}
\right), \,\,\, \ket{y-} = 2^{-1/2}
\left(
\begin{array}{c}
  1    \\
  -\mi \\
\end{array}
\right) \\
\ket{z+} &=& 
\left(
\begin{array}{c}
  1    \\
  0 \\
\end{array}
\right), \,\,\, \ket{z-} = 
\left(
\begin{array}{c}
  0    \\
  1 \\
\end{array}
\right)
\end{eqnarray}
The bras become row vectors; for example
\[
\bra{1} = (1, \,\,\,0), \,\,\, \bra{0}  = (0, \,\,\, 1)
\]
Notice in particular that
\[
\bra{y\pm} = 2^{-1/2}(\bra{1}\mp \mi \bra{0}) = 2^{-1/2} (1, \,\,\,\mp \mi)
\]

The outer product provides the projectors
\begin{eqnarray}
\mathcal{P}_{x+} &=&\frac{1}{2}
\left(
\begin{array}{cc}
  1  & 1  \\
  1 & 1\\
\end{array}
\right), \,\,\, \mathcal{P}_{x-}= \frac{1}{2}
\left(
\begin{array}{cc}
  1 & -1   \\
  -1 & 1\\
\end{array}
\right) \label{px} \\
\mathcal{P}_{y+} &=& \frac{1}{2}
\left(
\begin{array}{cc}
  1 & -\mi   \\
  \mi & 1 \\
\end{array}
\right), \,\,\, \mathcal{P}_{y-} = \frac{1}{2}
\left(
\begin{array}{cc}
  1 & \mi   \\
  -\mi  &1\\
\end{array}
\right) \label{py}\\
\mathcal{P}_{z+} &=& 
\left(
\begin{array}{cc}
  1 & 0   \\
  0 & 0\\
\end{array}
\right), \,\,\, \mathcal{P}_{z-} = 
\left(
\begin{array}{cc}
 0 &0    \\
  0 &1 \\
\end{array}
\right)
\end{eqnarray}
Note that, for each $\xi$, $\mathcal{P}_{\xi +} + \mathcal{P}_{\xi -} = 1$, as required by (\ref{sp1}).

The spin matrices, or Pauli matrices, may be computed from (\ref{ap}). For example,
\[
\sigma_{x} = \mathcal{P}_{x+} - \mathcal{P}_{x-}
\]
The well-known results are
\begin{eqnarray}
\sigma_{x} &=& \left(
\begin{array}{cc}
  0 & 1   \\
  1 & 0\\
\end{array}
\right),  \\
\sigma_{y} &=& \left(
\begin{array}{cc}
  0 & -\mi   \\
  \mi & 0\\
\end{array}
\right) \\
\sigma_{z} &=& \left(
\begin{array}{cc}
  1 & 0   \\
  0 & -1\\
\end{array}
\right) 
\end{eqnarray}

It is also worth noting that
\begin{eqnarray}
\sigma_{x}\ket{1} &=& \ket{0},\,\,\, \sigma_{x}\ket{0} = \ket{1} \\
\sigma_{y}\ket{1} &=& \mi \ket{0}, \,\,\, \sigma_{y}\ket{0} = -\mi \ket{1}
\end{eqnarray}
%
%

\subsection{Spin 1}
Considering the three-dimensional spin-$1$ space, we denote the spin operator by $S_{\xi}$ and the eigenstates by $\ket{\xi \lambda}$, where $\xi = (x,y,z)$ and $\lambda = (-1,0,1)$.  Thus, for example
\[
S_{x}\ket{x+} = \ket{x+}
\]
We choose the three basis kets $(\ket{z -}, \ket{z 0},\ket{z +})$ and find
\begin{eqnarray}
S_{x}  = 2^{-1/2}\left(
\begin{array}{ccc}
0 & 1& 0 \\
1 & 0 & 1 \\
0 & 1 & 0 \\
\end{array} \right) \\
S_{y}  = \mi 2^{-1/2}\left(
\begin{array}{ccc}
0 & -1& 0 \\
1 & 0 & -1 \\
0 & 1 & 0 \\
\end{array} \right) \\
S_{z} = \left(
\begin{array}{ccc}
1 & 0& 0 \\
0 & 0 & 0 \\
0 & 0 & -1 \\
\end{array} \right)
\end{eqnarray}

We then compute
\begin{eqnarray}
S_{x}^{2}  = \frac{1}{2}\left(
\begin{array}{ccc}
1 & 0& 1\\
0 & 2 & 0 \\
1 & 0 & 1 \\
\end{array} \right)\\
S_{y}^{2}  = \frac{1}{2}\left(
\begin{array}{ccc}
1 & 0& -1 \\
0 & 2 & 0 \\
-1 & 0 & 1 \\
\end{array} \right)\\
S_{z}^{2} = \left(
\begin{array}{ccc}
1 & 0& 0 \\
0 & 0 & 0 \\
0 & 0 & 1 \\
\end{array} \right)
\end{eqnarray}

The eigenvalues $\lambda$ of the $S_{i}^{2}$ are easily computed.  In all cases we find
\[
\lambda = 0,1
\]
where $\lambda = 1$ is degenerate.  

It is perhaps surprising but easily verified that all these operators commute:
\begin{equation}
\label{coms}
[S_{i}^{2},S_{j}^{2}] = 0, \text{for}\,\, (i,j)  = 1,2,3
\end{equation}
Therefore the corresponding eigenvalues can be simultaneously observed.  Their sum must satisfy
\[
S_{x}^{2} + S_{y}^{2} + S_{z}^{2}  = S^{2} = s(s+1) = 2
\]
That is, any measurement of all three must yield the values $(1,1,0)$ in some order.

Commutation implies the existence of simultaneous eigenstates.  These can be chosen to be
\begin{eqnarray}
\left(
\begin{array}{c}
2^{-1/2}   \\
0 \\
2^{-1/2} 
 \end{array}
\right), \,\,\, \left(
\begin{array}{c}
0    \\
1 \\
0
 \end{array}
\right)
\left(
\begin{array}{c}
-2^{-1/2}   \\
0 \\
2^{-1/2} 
 \end{array}
\right)
\end{eqnarray}

Finally we display a relation between the squared spin operators and the projectors
\[
P_{x+} = \ket{x+}\bra{x+}, \ldots, P_{z-} = \ket{z-}\bra{z-}
\]
The eigenkets are easily found; it suffices to display
\begin{eqnarray}
P_{x+}  = \frac{1}{4}\left(
\begin{array}{ccc}
1 & \sqrt{2}& 1\\
\sqrt{2} & 2 & \sqrt{2} \\
1 & \sqrt{2} & 1 \\
\end{array} \right)\\
P_{x0} = \frac{1}{2}\left(
\begin{array}{ccc}
1 & 0& -1 \\
0 & 0& 0 \\
-1 & 0 & 1 \\
\end{array} \right)\\
P_{x-} = \frac{1}{4}\left(
\begin{array}{ccc}
1 & -\sqrt{2}& 1 \\
-\sqrt{2} & 2 & -\sqrt{2} \\
1 & -\sqrt{2} & 1 \\
\end{array} \right)
\end{eqnarray}
These operators can be seen to satisfy $P_{x+}  + P_{x0} + P_{x-} = \bI$. (They are symmetric because they are real and Hermitian.) We do not display the remaining projectors, but note the identities
\begin{equation}
\label{sip}
S_{x}^{2} = \bI - P_{x0}, \,\, S_{y}^{2} = \bI - P_{y0}, \,\, S_{z}^{2} = \bI - P_{z0},
\end{equation}
The zero-spin states are orthogonal: $\braket{x0|y0} = \braket{x0|z0} = \braket{y0|z0}  = 0$, so products of the zero-spin projectors all vanish, and  and the $P_{\xi 0}$ commute. (This does not hold for non-zero spin.). Therefore (\ref{sip}) implies (\ref{coms}). 

For a simple derivation of (\ref{sip}) let $\xi$ denote $x,y,$ or $z$, so that the eigenvalue equation for $S_{\xi}$ is
\[
S_{\xi}\ket{\xi \lambda} = \lambda \ket{\xi \lambda} 
\]
where $\lambda = -1,0,1$. Then
\[
S_{\xi}^{2}\ket{\xi \lambda} = \lambda^{2} \ket{\xi \lambda} 
\]
Now consider
\[
(S_{\xi}^{2} +P_{\xi 0})\ket{\xi \lambda} = (\lambda^{2} + \delta_{\lambda 0}) \ket{\xi \lambda} 
\]
By considering the two cases $\lambda^{2} = 1,0$ we find that
\[
(S_{\xi}^{2} +P_{\xi 0})\ket{\xi \lambda} = \ket{\xi \lambda} 
\]
whence $S_{\xi}^{2} + P_{\xi 0} = \bI$.

\paragraph{Non-uniqueness of partitions}
From completeness, as in (\ref{sp1}), we know that
\[
\bI = P_{x0} + P_{x+} + P_{x-}
\]
while from (\ref{sip}) it is easy to show that
\begin{equation}
\label{sip2}
\bI = P_{x0} + P_{y0} + P_{z0}
\end{equation}
Indeed we see that $\{\ket{x0}, \ket{y0}, \ket{z0}\}$ form a CO set, so (\ref{sip2}) also follows from completeness.  

We conclude that distinct partitions of unity, each containing the same projector (in this case $P_{x0}$), can coexist.  The experimental significance concerns \emph{textuality}: when some quantity is measured, can the result depend upon what other quantities are simultaneously measured?

\setcounter{chapter}{2}
\chapter{The density operator}\label{Ch2DEO}

\section{Density operator}
\subsection{Definitions}
We consider an operator $\bm{\rho}$ that is expressed, in terms of some CO set of states $\chi_{n}$, as a linear combination of dyads:
\begin{equation}
\label{defr}
\bro = \sum_{m n}\rho_{mn}\ket{\chi_{m}}\bra{\chi_{n}}
\end{equation}
This operator is a \emph{density operator} iff it satisfies three conditions: \begin{enumerate}
\item It is Hermitian: $\rho_{m n} = \rho^{*}_{n m}$.  (Recall that the $\rho_{n m}$ are simply the components of $\bm{\rho}$ in the $\chi_{\mu}$ basis.)
\item It has unit trace: $\sum_{n}\rho_{nn} = 1$.  (The operator trace in an infinite-dimensional space can be problematic; we do not address this complication.)
\item  It is a positive operator: $\bra{\psi}\rho \ket{\psi} \geq 0$ for any state $\psi$.
\end{enumerate}
In an important special case the sum reduces to a single term: $\bro = \ket{\psi} \bra{\psi}$.  Then the operator is said to describe a \emph{pure state}.  The general case with more than one term describes a \emph{mixed state}. 

The trace of the \emph{squared} density operator for a mixed state is always less than unity, as can be seen from the diagonal form:
\[
\bro^{2}  = \sum_{mn}p_{m}p_{n}\mathcal{P}_{m}\mathcal{P}_{n} = \sum_{n}p_{n}^{2}\mathcal{P}_{n}
\]
so, in view of (\ref{trp1}),
\[
\text{Tr}(\bro^{2})  = \sum_{n}p_{n}^{2} < \sum_{n}p_{n} = 1
\]
since $p_{n}<1$ for each $n$.  However the density operator for a pure state is a projector and therefore the trace of its square is unity.

In summary, neither pure states nor mixed states are  vectors (or rays) in a Hilbert space; rather each corresponds to an operator.  A pure state is represented by a projector; we next observe that a mixed state can be represented by a linear combination of projectors.

\subsection{Proper density operators}\label{subpro}
The proper form of a density operator, often called the \emph{proper density operator}, is diagonal:
\begin{equation}
\label{defr2}
\bm{\rho} = \sum_{m}p_{m}\ket{\phi_{m}} \bra{\phi_{m}} = \sum_{m}p_{m}\mathcal{P}_{m}
\end{equation}
Thus the proper density operator is a sum of projectors. The proper form has a simple interpretation: it describes a system whose quantum state is known only to be one of the states $\psi_{m}$, with probability $p_{m}$.  

The set of states $\phi_{m}$ might form a continuum:  $\phi_{m} \rightarrow \phi(x)$.  In that case (\ref{defr2}) becomes
\begin{equation}
\label{rc}
\bro = \int \dif x \, p(x) \ket{\phi(x)} \bra{\phi(x)}
\end{equation}
 
Since the matrix $\rho_{mn}$ can always be diagonalized, every density operator has a proper expression in terms of its eigenstates. However proper expression does not require expression in terms of eigenstates. Equation (\ref{defr2}) assumes that the kets $\ket{\phi_{m}}$ are normalized, 
\[
\braket{\phi_{m}|\phi_{m}} = 1
\]
but it does \emph{not} require that they are orthogonal: we allow $\braket{\phi_{m}|\phi_{n}} \neq 0$. Thus the $\ket{\phi_{m}}$ need not belong to a CO set.  In the special case when the  $\ket{\phi_{m}}$  \emph{do} belong to a CO set, it can be seen that this set is simply the eigenkets of $\bro$.  That is,\begin{quote} In any proper expression for $\bro$, either each term projects onto eigenstates of $\bro$, or the projectors are not orthogonal.\end{quote}

The correspondence between a quantum system and its proper density operator is not simple, for two reasons.  \begin{enumerate}  \item The assignment of probabilities is fully objective only when ``state preparation'' is unambiguously defined.  As remarked in subsection \ref{density}, this definition is not problematic in the standard context of quantum states:  one simply requires measurement of a complete set of commuting observables.  But the standard requirement makes no sense in the density operator context, and a replacement is not obvious.  Can different experimental configurations, poised to act on a single quantum system, correspond to different density operators? (Bohr's response to this issue was to include any measuring devices in the definition of the system.\cite{bohr1958})
 \item We will see in subsection \ref{nonuni}, below, that the proper expression of any density operator is not unique: a given operator will have many proper expressions, each involving a distinct set of states. Only its expression in terms of CO states, which are necessarily eigenstates, is unique (up to degeneracy). \end{enumerate} 

For  a proper operator the unit trace requirement becomes
\[
\sum_{m}p_{m} = 1
\]
and positivity is expressed by $p_{m} \geq 0$.  That is,
\[
\braket{v|\rho|v}  = \sum_{i}p_{i}\braket{v|\psi_{i}}\braket{\psi_{i}|v} = \sum_{i}p_{i}|v_{i}|^{2} \geq 0
\]
where $v_{i} = \braket{\psi_{i}|v}$.


Although the $\ket{\phi_{m}}$ need not be orthogonal, we can express them in terms of a CO set, denoted here by $\ket{\chi_{m}}$:
\[
\ket{\phi_{m}} = \sum_{k}a_{mk}\ket{\chi_{m}}, \,\,\, \bra{\phi_{m}} = \sum_{k}a^{*}_{mk}\bra{\chi_{m}}
\]
The normality condition $\braket{\phi_{m}|\phi_{m}} = 1$ then implies
\begin{equation}
\label{saa}
\sum_{k}a_{mk}a^{*}_{mk} = \sum_{k}|a_{mk}|^{2} = 1
\end{equation}
for each $m$.  Considering the $a_{mk}$ as elements of a 
matrix $\ba$, we see that the squared magnitudes of each matrix row must sum to unity.

We next express (\ref{defr2}) in terms of the $\ket{\chi_{m}}$, to obtain the improper representation
\[
\bro = \sum_{mn}\rho_{mn}\ket{\chi_{m}}\bra{\chi_{n}}
\]
with
\begin{equation}
\label{rpaa}
\rho_{mn} = \sum_{k}\,p_{k}a_{km}a^{\dagger}_{nk}
\end{equation}
Introducing $\hat{a}_{km} \equiv \sqrt{p_{k}}\, a_{km}$, we can write $\bro$ as the matrix product
\begin{equation}
\label{raa}
\bro = \hat{\ba}^{\dagger}\cdot \hat{\ba}
\end{equation}
It is easily verified, using (\ref{saa}), that the trace condition $\text{Tr}(\bro) = 1$ is satisfied. 

\subsection{Non-uniqueness} \label{nonuni}
The proper representation (\ref{defr2}) of $\bro$ is not unique. In general there will exist various sets of states---various mixtures---that represent, properly, a given density operator.  The simplest case pertains to a degenerate operator---a $\bro$ with multiple eigenvectors. In a very simple case we have the operator (in the notation of Section \ref{spinal})
\[
\bro_{*} = \frac{1}{2}(\ket{1}\bra{1} + \ket{0}\bra{0})
\]
This operator (in fact a multiple of the identity) is highly degenerate, and could just as well be written
\[
\bro_{*} = \frac{1}{2}(\ket{x+}\bra{x+} + \ket{x-}\bra{x-})
\]
The formulae of Section \ref{spinal} quickly show that these expressions refer to the same operator.  Yet the preparations of the two densities are clearly distinct, employing different orientations of a Stern-Gerlach device.  From the perspective of quantum states, one has two quite different situations---a difference that the density operator does not reflect. (It can be argued that macro-systems are unlikely to show degeneracy \cite{barandes2019}).

A more important instance of non-uniqueness does not require degeneracy.  One considers states $\psi_{m}$ that are normalized
\[
\braket{\psi_{m} | \psi_{m}} = 1
\]
but \emph{not} orthogonal.  Then the proper expression
\begin{equation}
\label{ralt}
\bro = \sum_{n}r_{n} \ket{\psi_{n}}\bra{\psi_{n}}
\end{equation}
can represent the operator of (\ref{defr2}).

Non-uniqueness of the proper expression is important. It implies that the statement ``The state of the system is one of the states $\phi_{n}$ with probabilities $p_{n}$'' is a much stronger statement than (\ref{defr2}), and not equivalent to it.  For the given $\bro$ one could just as accurately say ``The state of the system is one of the states $\psi_{n}$ with probabilities $r_{n}$''.

To understand the relation between the two representations, we expand the $\ket{\psi_{n}}$ in terms of the same CO set used for the $\ket{\phi_{n}}$:
\begin{equation}
\label{defpsim}
\ket{\psi_{k}} = \sum_{m}b_{km}\ket{\chi_{m}}, \,\,\, \bra{\psi_{k}} = \sum_{m}b^{*}_{km}\bra{\chi_{m}}
\end{equation} 
The non-uniqueness of the proper expression is then expressed by
\begin{equation}
\label{punch0}
\sum_{k}p_{k}a_{km}a^{*}_{kn} = \sum_{k}r_{k}b_{km}b^{*}_{kn}
\end{equation}
In term of the modified matrices $\hat{a}_{km} = \sqrt{p_{k}}\, a_{km}$ and $\hat{b}_{km} = \sqrt{r_{k}} \, b_{km}$, (\ref{punch0}) becomes the matrix equation
\begin{equation}
\label{punch7}
\hat{\ba}^{\dagger}\cdot \hat{\ba} =  \hat{\bb}^{\dagger}\cdot \hat{\bb}
\end{equation}
which requires
\[
\hat{\bb} = \bu \cdot \hat{\ba}
\]
where $\bu$ is an arbitrary unitary matrix, $\bu^{\dagger}  \cdot\bu = \bI$.  The point is that $\bro$ is invariant under the transformation $\hat{\ba}  \rightarrow \bu \cdot \hat{\ba}$, but this transformation changes the states in the proper expression.  Alternative discussions of non-uniqueness can be found in \cite{gisin1989} and \cite{hughston1993}.

To summarize,  given a proper density operator $\bro$ projecting onto states $\ket{\phi_{m}}$, we choose a CO set to compute the coefficients  $a_{mk}$  and the matrix $\hat{\ba}$.  To find a distinct but also proper expression for $\bro$, we choose an arbitrary unitary matrix $\bu$ and let $\hat{\bb} = \bu \cdot \hat{\ba}$.  We next compute the $r_{m}$ by imposing (\ref{saa}): $r_{m}= \sum_{k}|\hat{b}_{mk}|^{2}$.  The resulting coefficients $b_{mk}$ provide, through (\ref{ralt}) and (\ref{defpsim}), the desired alternative expression.

\paragraph{Example}
We consider the two-dimensional spin-$1/2$ system with the pro\-per density operator
\begin{equation}
\label{npr1}
\bro = \frac{1}{2}(\mathcal{P}_{x+} + \mathcal{P}_{y+})
\end{equation}

That is, we know that the system spin is in either the positive-$x$ or the positive-$y$ direction, with equal probabilities.  Since the two states are not orthogonal, $\braket{x+|y+} = (1/2)(1 + \mi)$, the projectors are not orthogonal. In terms of the standard basis states $\ket{0}$ and $\ket{1}$, we find that $\bro$ is represented by the (improper) matrix
\begin{equation}
\label{rmm}
\bro = \frac{1}{2}  \begin{pmatrix}
 1   &  \frac{1}{2}  (1+\mi)  \\
 \frac{1}{2}  (1-\mi)     &  1
\end{pmatrix}
\end{equation}

A  proper expression for $\bro$ which is distinct from (\ref{npr1}) uses the states
\[
\ket{\chi_{1}} =3^{-1/2}(\ket{x+} + \ket{y+}), \,\,\, \ket{\chi_{2}} = \ket{x+} - \ket{y+}
\]
These are normalized, $\braket{\chi_{i} | \chi_{i}} = 1$, but also non-orthogonal.  Using
\[
\ket{x+} = \frac{1}{2}(\sqrt{3} \ket{\chi_{1}} + \ket{\chi_{2}}), \,\, \ket{y+} = \frac{1}{2}(\sqrt{3} \ket{\chi_{1}} - \ket{\chi_{2}})
\]
we find that our density operator has the alternative form
\[
\bro = \frac{3}{4}\mathcal{P}_{\chi_{1}} + \frac{1}{4}\mathcal{P}_{\chi_{2}}
\]
Straightforward algebra confirms that this expression has the same matrix, (\ref{rmm}).

The representation matrices are 
\begin{eqnarray*}
\ba &=& \begin{pmatrix}
 2^{-1/2}     & 2^{-1/2}   \\
      2^{-1/2}  &   2^{-1/2} \mi
\end{pmatrix} \\
\bb &= &\begin{pmatrix}
   \sqrt{2/3}   & 6^{-1/2} (1+\mi)  \\
    0  &  2^{-1/2} (1 - \mi)
\end{pmatrix}
\end{eqnarray*}
It is clear that (\ref{saa}) is satisfied.  The ``hatted'' matrices are
\begin{eqnarray*}
\hat{\ba} &=& \frac{1}{2} \begin{pmatrix}
 1 & 1   \\
      1  &    \mi
\end{pmatrix} \\
\hat{\bb} &= &\frac{1}{\sqrt{2}}\begin{pmatrix}
1  & \frac{1}{2}(1+\mi)  \\
    0  &  \frac{1}{2} (1 - \mi)
\end{pmatrix}
\end{eqnarray*}
corresponding to the unitary matrix
\[
\bu = 2^{-1/2}\begin{pmatrix}
   1   &  1  \\
    1  &  -1
\end{pmatrix}
\]
Simple matrix multiplication confirms the relations
\[
\hat{\ba}^{\dagger}\cdot \hat{\ba} = \hat{\bb}^{\dagger}\cdot \hat{\bb} = \bro
\]

\subsection{Expectation values}
Consider the mixed state
\[
\bro = \sum_{m}p_{m}\mathcal{P}_{m}, \,\, \mathcal{P}_{m} = \ket{\psi_{m}}\bra{\psi_{m}}
\]
where the states $\psi_{m}$  are in general non-othogonal,  
The expected value of an operator $\bK$ in the state $\ket{\psi_{m}}$ is 
 \[
\braket{K}_{m} = \braket{\psi_{m}|\bK |\psi_{m}}
\]
so the expectation value of $K$ in the mixed state is
\begin{equation}
\label{defav}
\braket{K} = \sum_{m}p_{m}\braket{\psi_{m}|\bK |\psi_{m}}
\end{equation}
For example the expectation of a projection operator, $\mathcal{P}_{\chi} \equiv \ket{\chi}\bra{\chi}$ is
\begin{equation}
\label{avp}
\braket{\mathcal{P}_{\chi}} = \sum_{n}p_{m}\braket{\psi_{n}|\chi}\braket{\chi|\psi_{m}} = \sum_{m}p_{m}|\braket{\psi_{m}|\chi}|^{2}
\end{equation}
It is easily seen that this formula is equivalent to
\[
\braket{\mathcal{P}_{\chi}} = \braket{\chi | \bro |\chi}  = \braket{\bro}_{\chi}
\]

As an Hermitian operator, $\mathcal{P}_{\chi}$ has a CO set of eigenstates $\chi_{m}$. Denoting the projector for eigenstate $\chi_{k}$ by $\mathcal{P}_{k}$, we see that
\[
\braket{\mathcal{P}_{k}}_{m} = \delta_{km}
\]
so that (\ref{defav}) implies
\[
\braket{\mathcal{P}_{k}} = \sum_{m}p_{m} \delta_{km} = p_{k}
\]
In terms of the $\chi_{m}$ basis, the density operator can be expressed as 
\begin{equation}
\label{pp}
\bm{\rho} = \sum_{k}\braket{\mathcal{P}_{k}}\mathcal{P}_{k}
\end{equation}

\paragraph{Trace formula}
A general and useful formula for the average of an observable $\bK$ is
\begin{equation}
\label{tr}
\braket{K}  = \text{Tr}(\bm{\rho} \bK)
\end{equation}
(We recall here that $\text{Tr}(\bA  \bB) = \text{Tr}(\bB  \bA)$.) Indeed, taking the $\psi_{m}$ to form a CO set, and starting with the diagonal representation (\ref{defr2}), we find
\begin{eqnarray}
 \text{Tr}(\bm{\rho} \bK) \equiv \sum_{m}\braket{\psi_{m}| \bm{\rho} \bK |\psi_{m}} &=&  \sum_{mn} \delta_{mn} P_{n} \braket{\psi_{n}|\bK |\psi_{m}} \nonumber\\
 & = & \sum_{m} P_{m }\braket{K}_{m} \label{ppk}
\end{eqnarray}
as in (\ref{tr}). Since the trace is independent of basis, this trace formula applies for any CO basis.  

Recalling that all the $P_{n}$ are positive, we see an important consequence of (\ref{ppk}): if $\text{Tr}(\bro \bK) = 0$ for a Hermitian operator $\bK$ and any density operator $\bro$, then we must have $\bK = 0$.

Since the observable $\bK$ is Hermitian, it has eigenkets $\ket{\psi_{\ell}}$ and eigenvalues $k_{\ell}$, whence the expression
\[
\bK = \sum_{\ell}k_{\ell}\ket{\psi_{\ell}}\bra{\psi_{\ell}}
\]
In the same basis, the density operator is 
\[
\bro = \sum_{mn}\rho_{mn}\ket{\psi_{m}}\bra{\psi_{n}}
\]
Therefore
\[
\bro \bK = \sum_{mn}k_{n}\rho_{mn}\ket{\psi_{m}}\bra{\psi_{n}}
\]
and we easily find
\begin{equation}
\label{tbk}
\text{Tr}(\bro\bK ) = \sum_{n}k_{n}\rho_{nn}
\end{equation}

Consider next the proper density operator,
\[
\bm{\rho} = \sum_{m} p_{m}\ket{\phi_{m}} \bra{\phi_{m}}
\]
where the $\phi_{m}$ are \emph{not} orthogonal.  Now the trace must be computed in terms of a CO basis, yet the naive formula  
\begin{equation}
\text{Tr}(\bro \bK) = \braket{K} = \sum_{m}p_{m}\braket{\phi_{m}|\bK |\phi_{m}}
\label{nk}
\end{equation}
pertains.  To verify this fact, we expand the $\phi_{m}$ in terms of the CO set $\psi_{m}$,
\[
\ket{\phi_{m}} = \sum_{k} C_{mk}\ket{\psi_{k}}
\]
and find
\[
\bro = \sum_{mn}\hat{\rho}_{mn}\ket{\psi_{m}}\bra{\psi_{n}}
\]
where
\[
\hat{\rho}_{mn} = \sum_{\ell}p_{\ell}C^{*}_{\ell m}C_{\ell n}
\]
Thus (\ref{tbk}) yields
\begin{equation}
\label{tbk2}
\text{Tr}(\bro \bK) = \sum_{n}k_{n} \sum_{\ell}p_{\ell}|C_{\ell n}|^{2}
\end{equation}
On the other hand
\[
\braket{K}_{m} = \sum_{n}k_{n}|C_{mn}|^{2}
\]
so the definition, (\ref{defav}) gives
\[
\braket{K} = \sum_{m n}p_{m}k_{n}|C_{mn}|^{2}
\]
in agreement with (\ref{tbk2}).

\section{Dynamics of density operators}
\subsection{Isolated system}
The proper density operator
\begin{equation}
\label{r7}
\bm{\rho} = \sum_{m}p_{m}\ket{\phi_{m}} \bra{\phi_{m}} = \sum_{m}p_{m}\mathcal{P}_{m}
\end{equation}
describes a system that is known to be in one of the states $\phi_{m}$, with probability $p_{m}$.  If the system remains isolated, so that no new information is obtained, these probabilities will not change.  But the projectors do change, by Schr\"odinger evolution, as discussed previously.  It follows from (\ref{cp}) that, for an isolated system,
\begin{equation}
\label{rt}
\p_{t} \bm{\rho} = -\mi [H, \bm{\rho}]
\end{equation}
It is often convenient to express this dynamic in terms of the solution 
\begin{equation}
\label{rt2}
\bm{\rho}(t) = \me^{-\mi tH} \bm{\rho}(0) \me^{\mi tH}
\end{equation}
This evolution is evidently unitary.

\subsection{Measured system}
Consider again a mixed state initially described by the operator $\bro(0)$ having the form of (\ref{r7}). A measurement of the system, corresponding to some operator with eigenstates $\chi_{m}$, imposes a new set of projectors 
\[
\mathcal{R}_{m} = \ket{\chi_{m}}\bra{\chi_{m}}
\]
and a new density operator, $\bm{\rho}(0) \rightarrow \bro(t)$. Note that the states $\chi_{m}$, corresponding to an observable (Hermitian operator), can be assumed to form a CO set. Since the measurement is performed on the original mixed state, the probability $R_{m} = \braket{\mathcal{R}_{m}}$ is given by (\ref{avp}):
\[
\braket{\mathcal{R}_{m}} = \sum_{n}p_{n}|\braket{\chi_{n}|\phi_{m}}|^{2}
\]
From (\ref{pp}) we have
\begin{equation}
\label{md}
\bro(t)  = \sum_{m}\braket{\mathcal{R}_{m}}\mathcal{R}_{m}  = \sum_{mn}p_{n}\,|\braket{\chi_{n}|\phi_{m}}|^{2}\,\mathcal{R}_{m}
\end{equation}
The trace formula (\ref{tr}) provides an alternative expression,
\begin{equation}
\label{r3}
\bro(t) = \sum_{m}\mathcal{R}_{m} \text{Tr}(\bm{\rho}(0)\mathcal{R}_{m})
\end{equation}
which further simplifies:
\begin{equation}
\label{r4}
\bro(t) = \sum_{m}\mathcal{R}_{m} \bm{\rho}(0)\mathcal{R}_{m}
\end{equation}

To show that (\ref{r3}) and (\ref{r4}) are equivalent, we express the former as
\begin{eqnarray*}
 \sum_{m}\mathcal{R}_{m} \text{Tr}(\bm{\rho}(0)\mathcal{R}_{m}) &=& \sum_{m k}\mathcal{R}_{m} \bra{\chi_{k}}\bro(0)\ket{\chi_{m}}\braket{\chi_{m}|\chi_{k}} \\
 &=& \sum_{m}\ket{\chi_{m}}\bra{\chi_{m}} \bra{\chi_{m}}\bro(0)\ket{\chi_{m}}
\end{eqnarray*}
or, after rearranging a bra,
\[
 \sum_{m}\mathcal{R}_{m} \text{Tr}(\bm{\rho}(0)\mathcal{R}_{m}) = \sum_{m}\ket{\chi_{m}} \bra{\chi_{m}}\bro_{0}\ket{\chi_{m}}\bra{\chi_{m}} =  \sum_{m}\mathcal{R}_{m} \bm{\rho}(0)\mathcal{R}_{m}
\]
as in (\ref{r4}).

To check the trace condition, we recall that the trace of a triple product allows rearrangement, so that
$\text{Tr}(\mathcal{R}_{m} \bm{\rho}(0)\mathcal{R}_{m}) = \text{Tr}(\bm{\rho}(0))\mathcal{R}^{2}_{m})  = \text{Tr}(\bm{\rho}(0)\mathcal{R}_{m})$.  Therefore
\[
\text{Tr}(\bro(t)) = \sum_{m}\text{Tr}(\bro(0))\mathcal{R}_{m} = \text{Tr}(\bro(0)) \sum_{m}\mathcal{R}_{m} = \text{Tr}(\bro(0))
\]

\subsection{Quantum evolutions}\label{qevol}
Equations (\ref{rt2}) and (\ref{md}) display the notoriously disparate evolution laws of quantum theory:
\begin{eqnarray}
\bm{\rho}(t) &=& \me^{-\mi tH} \bm{\rho}(0) \me^{\mi tH} \label{qt1}\\
\bm{\rho}(t) &=& \sum_{m}\mathcal{R}_{m} \bm{\rho}(0)\mathcal{R}_{m}\label{qt2}
\end{eqnarray}
where, as above, $\mathcal{R}_{m}$ is the projector corresponding to the measured observable.  This disparity is associated with ``collapse of the wave function.''  Note in particular that the projector $\mathcal{R}_{m}$ is not a unitary operator.  The Lindblad equation, derived in Chapter \ref{Ch6LIN}, can be considered a bridge between (\ref{qt1}) and (\ref{qt2}). 

Collapse of the wave function is conventionally viewed in terms of the quantum state, rather than the density operator.  Thus suppose the initial state of some system is $\Psi_{0}$, and the post-measurement state is one of two macroscopically distinct states $\ket{\psi_{1}}$ or $\ket{\psi_{2}}$, with $\braket{\psi_{1}|\psi_{2}} = 0$. Then the measurement process cannot be unitary: unitary evolution would give $\ket{\psi_{i}}  = U\ket{\Psi(0)}, \,\,i = 1,2$, and therefore $\braket{\psi_{1}| \psi_{2}} = \braket{\Psi(0) U^{\dagger} U \Psi(0)} = 1$.  

The collapse process can be decomposed into two steps: (1) the interaction of system and measuring apparatus, producing the state described by (\ref{qt2}); (2) the observation of the result, producing a pure state, with $\bro$ given by a single projector.  Because the observation step leads to various philosophical problems (must the observer be human? conscious?\cite{wigner1995}) it is commonly ignored: measurement of the system is assumed to produce the state (\ref{qt2}). It is significant that, even without the observation step, measurement causes the density operator to change in a non-unitary, non-Hamiltonian way.

\setcounter{chapter}{3}
\chapter{Bipartite systems}\label{Ch3BIP}

\section{Bipartite systems}\label{bs}
\subsection{States}\label{states}

Many systems are conveniently decomposed into two subsystems.  The resulting Hilbert space is then viewed as the tensor product of two distinguishable Hilbert spaces:
\[
\mathcal{H} = \mathcal{H}_{a}\otimes \mathcal{H}_{b}
\]
(Decomposition into more than two subspaces is not discussed here.) For example, $\mathcal{H}_{a}$ might be an atomic system and $\mathcal{H}_{b}$ a macroscopic measuring device; or both subspaces might be microscopic, but separated spatially.  We adopt the convention of denoting states in $\mathcal{H}_{a}$ ($\mathcal{H}_{b}$) by $\phi$ ($\chi$).  If the states $\phi_{n}$ and $\chi_{n}$ are CO in their respective Hilbert spaces, then the states $\ket{\Psi_{mn}} = \ket{\phi_{m}}\otimes \ket{\chi_{n}}$ are CO in $\mathcal{H}$.   Often we use an abbreviated notation, omitting the symbol $\otimes$.  Notice that \begin{enumerate} \item The first (second) member of the double index refers to the space $\mathcal{H}_{a}$ ($\mathcal{H}_{b}$).  \item Any expression involving $\braket{\phi | \chi }$ is meaningless, since the two states are in different Hilbert spaces. For the same reason such kets commute: $\ket{\phi} \ket{\chi} = \ket{\chi} \ket{\phi}$.\end{enumerate}

\paragraph{Product states}
The simplest states in $\mathcal{H}$, called product states, have the form
\[
\ket{\Psi} = \ket{\phi}\otimes \ket{\chi} 
\]
We expand
\[
\ket{\phi} = \sum_{m}A_{m}\ket{\phi_{m}}, \,\,\, \ket{\chi} = \sum_{m}B_{m}\ket{\chi_{m}}
\]
to write the product state as
\begin{eqnarray}
\ket{\Psi} &=& \sum_{m}A_{m}\ket{\phi_{m}}\sum_{n}B_{n}\ket{\chi_{n}} \label{f1} \\
&=& \sum_{\bem}C_{\bem} \ket{\psi_{\bem}} \label{f2}
\end{eqnarray}
where $\bem = (mn)$, $C_{\bem}  = A_{m}B_{n}$ and
\begin{equation}
\label{defpm}
\ket{\psi_{\bem}} = \ket{\phi_{m}}\ket{\chi_{n}}
\end{equation}
Note that product states  in $\mathcal{H}$, and therefore all states need double indices.

Any state in $\mathcal{H}$ has an expression of the form (\ref{f2}); what is special about product states is that $C_{\bem}$ factors: \begin{equation}
\label{prod}
C_{\bem} = A_{m}B_{n}
\end{equation}
This factorization is necessary and sufficient for the state to be a product state. When there are no such factors, the bipartite state is said to be \emph{entangled}. Entanglement is essential to quantum measurement:  a measurement occurs when the state of some quantum system is entangled with the state of a measuring device.

\paragraph{Entangled states}
 The simplest example of an entangled state is
\begin{equation}
\label{tan}
\ket{\Psi_{e}} = 2^{-1/2}(\ket{\phi^{1}}\ket{\chi^{1}} + \ket{\phi^{2}}\ket{\chi^{2}} )
\end{equation}
where $\ket{\phi}$ ($\ket{\chi}$) belongs to $\mathcal{H}^{a}$ ($\mathcal{H}^{b}$). We may expand
\[
\ket{\phi^{s}} = \sum_{m}A^{s}_{m}\ket{\phi_{m}}, \,\, \ket{\chi^{s}} = \sum_{m}B^{s}_{m}\ket{\chi_{m}}, \,\, s = 1,2
\]
and obtain (\ref{f2}), where now
\[
C_{\bem} = 2^{-1/2}(A^{1}_{m}B_{n}^{1} +  A_{m}^{2}B_{n}^{2})
\]

Entanglement is not always obvious.  Consider the states
\[
\ket{\phi} = A_{1}\ket{\phi_{1}} + A_{2}\ket{\phi_{2}}, \,\, \ket{\chi} = B_{1}\ket{\chi_{1}} + B_{2}\ket{\chi_{2}}
\]
The product state
\[
\ket{\psi} = A_{1}B_{1}\ket{\phi_{1}}\ket{\chi_{1}} + A_{1}B_{2}\ket{\phi_{2}}\ket{\chi_{1}} +A_{2}B_{1}\ket{\phi_{2}}\ket{\chi_{1}} +A_{2}B_{2}\ket{\phi_{2}}\ket{\chi_{2}} 
\]
might be mistaken for an entangled state. Moreover, if any term on the right-hand side is omitted, $\ket{\psi}$ becomes an entangled state.  The test of course is to express any state in the form (\ref{f2}) and then observe whether the coefficients factor.

\paragraph{Operators}
The components of an operator $\bV$ in $\mathcal{H}$ will have four indices:
\[
V_{k\ell mn} = V_{\bk \bem} = \bra{\psi_{\bk}}V\ket{\psi_{\bem}} = \bra{\phi_{k}}\bra{\chi_{\ell}} V \ket{\phi_{m}}\ket{\chi_{n}}
\]
where $\bk = (k,\ell)$. Certain operators $\bV$ may act only in $\mathcal{H}_{a}$, say, having no effect on the states of $\mathcal{H}_{b}$:
\[
\bV  = \bV^{a}\otimes \bI^{b}
\]
where $\bI^{b}$ is the identity operator in $\mathcal{H}_{b}$. Then
\[
\bV \ket{\Psi_{\bem}} = \ket{\chi_{n}}(\bV^{a}\ket{\phi_{m}}
\]
and
\[
\bV_{\bk\bem} = \braket{\psi_{\bk}| \bV^{a}|\psi_{\bem}} = V^{a}_{km}\delta_{\ell n}
\]

Some operators can be decomposed into a product, with factors parallel to those of the product space.  Thus an operator on $\mathcal{H}$ might have the form
\[
\bC = \bA \otimes \bB
\]
where the first (second) factor acts only on the space $\mathcal{H}_{a}$ ($\mathcal{H}_{b}$). In that case the components are computed from what is called the Kronecker product, a higher dimensional version of the tensorial vector product.  Thus each element of the matrix for $\bA$ multiplies the entire matrix for $\bB$:
\[
{\displaystyle \mathbf {A} \otimes \mathbf {B} ={\begin{bmatrix}a_{11}\mathbf {B} &\cdots &a_{1n}\mathbf {B} \\\vdots &\ddots &\vdots \\a_{m1}\mathbf {B} &\cdots &a_{mn}\mathbf {B} \end{bmatrix}},}
\]
The trace of a Kronecker product is the product of the two traces:
\[
\text{Tr}(\bA \otimes\bB) = \text{Tr} (\bA)\text{Tr}(\bB)
\]
A simple example is a measurement  that affects only subsystem $b$, corresponding to the operator
\[
\bV = \bI^{a}\otimes \bB
\]
Suppose $\bB$ is the projector $\mathcal{P}^{b}_{1} = \ket{\chi^{1}}\bra{\chi^{1}}$, and that it acts on the entangled state $\ket{\Psi_{e}}$ of (\ref{tan}). Then
\[
\bV = \bI^{a}\otimes\mathcal{P}^{b}_{1}
\]
and we have
\begin{equation}
\label{evp}
\bV \ket{\Psi_{e}} = 2^{-1/2}\ket{\Phi_{1}}\ket{\Psi_{1}}
\end{equation}
So subsystem $a$ is definitely in the state $\Phi_{1}$ after the measurement, even if the experimental devices of subsystems $a$ and $b$ avoided any interaction.  This circumstance is more than another example of wave function collapse; as we discuss elsewhere, it also involves issues of locality.


\subsection{Projectors and density operators}\label{prodsec}
The projector onto the product state $\ket{\psi}\ket{\chi}$,
\[
\mathcal{P}_{\ket{\psi}\ket{\chi}}  = \ket{\psi}\ket{\chi}\bra{\chi} \bra{\psi}
\]
satisfies
\[
\mathcal{P}_{\ket{\psi}\ket{\chi}} = \mathcal{P}_{\ket{\psi}}\mathcal{P}_{\ket{\chi}}
\]
where the first (second) factor acts in $\mathcal{H}_{a}$ ($\mathcal{H}_{b}$).  Bipartite projectors onto eigenstates have the form
\[
\mathcal{P}_{\bem} = \ket{\phi_{m}}\ket{\chi_{n}} \bra{\chi_{n}}\bra{\phi_{m}}
\]
and satisfy a version of (\ref{ppd}):
\[
\mathcal{P}_{\bk} \mathcal{P}_{\bem} = \delta_{\bk \bem}\mathcal{P}_{\bk}
\]
or, in expanded form,
\[
\mathcal{P}_{k\ell}\mathcal{P}_{mn} = \mathcal{P}_{k\ell}\delta_{km} \delta_{nl}
\]
Projectors are exceptional operators in needing only $2$, instead of $4$, indices.

Consider next the proper density operator,
\[
\bro = \sum_{\bem}p_{\bem}\ket{\Psi^{\bem}}\bra{\Psi^{\bem}}
\]
Here the $\ket{\Psi^{\bem}}$ are not eigenstates and need not be elements of a CO set.  Instead we have
\[
\ket{\Psi^{\bem}} = \sum_{\bem'}C^{\bem}_{\bem'}\ket{\psi_{\bem'}}
\]
and $\ket{\psi_{\bem'}} = \ket{\phi_{m'}}\ket{\chi_{n'}}$ is the eigenstate of (\ref{defpm}).  Then
\begin{equation}
\label{gr}
\bro = \sum_{\bem}P_{\bem}\sum_{\bem' \bem''}C^{\bem}_{\bem'}C^{\bem *}_{\bem''}\ket{\psi_{\bem''}}\bra{\psi_{\bem'}}
\end{equation}

In the unentangled case the density operator factors:  
\begin{equation}
\label{unt}
\bro = \bro^{a}\bro^{b}
\end{equation}
Indeed, we know that for a product state
\[
C^{\bem}_{\bem'} = C^{\bem}_{m'n'} = A^{\bem}_{m'}B^{\bem}_{n'}
\]
Substitution in to (\ref{gr}) yields the two improper density operators,
\[
\bro^{a} = \sum_{mm'm''}p_{m}A^{m}_{m'}A^{m *}_{m''}\ket{\phi_{m'}}\bra{\phi_{m''}}, \,\,  \bro^{b} = \sum_{nn'n''}p_{n}B^{n}_{n'}B^{n *}_{n''}\ket{\chi_{n'}}\bra{\chi_{n''}}
\]
That is
\[
\bro^{a} = \sum_{m' m''}\rho^{a}_{m'm''}\ket{\phi_{m'}}\bra{\phi_{m''}}, \,\,\bro^{b} = \sum_{m' m''}\rho^{b}_{m'm''}\ket{\chi_{m'}}\bra{\chi_{m''}}
\]
with 
\[
\rho^{a}_{m'm''} = \sum_{m}p_{m}A^{m}_{m'}A^{m *}_{m''}, \,\,\rho^{b}_{m'm''} = \sum_{m}p_{m}B^{m}_{m'}B^{m *}_{m''}
\]
More generally the systems are entangled: the coefficients to do not factor, so the density operator does not factor.
%

\subsection{Partial trace}
\paragraph{Projector trace}
Consider the projector $\mathcal{P}_{\Psi} = \ket{\Psi}\bra{\Psi}$ onto some state $\Psi$ in $\mathcal{H}$.  We can express the state as
\begin{equation}
\ket{\Psi} = \sum_{\bem}C_{\bem}\ket{\Psi_{\bem}}\label{pcm}
\end{equation}
Therefore
\begin{equation}
\label{p1}
\mathcal{P}_{\Psi} = \sum_{\bem \bk}C_{\bem}C^{*}_{\bk}\ket{\Psi_{\bem}}\bra{\Psi_{\bk}} 
\end{equation}
The \emph{partial trace} of this operator---the trace over the states $\chi_{m}$ of $\mathcal{H}_{b}$ only---is denoted by
\[
\text{Tr}^{b}(\mathcal{P}_{\Psi}) = \sum_{s}\braket{\chi_{s} |\mathcal{P}_{\Psi} |\chi_{s}}
\]
Using the orthonormality of the $\chi_{m}$, we find
\begin{equation}
\label{trp}
\text{Tr}^{b}(\mathcal{P}_{\Psi}) = \sum_{mnk}C_{mn}C^{*}_{kn}\ket{\phi_{m}}\bra{\phi_{k}}
\end{equation}
In the product case, 
\[
\ket{\Psi} = \ket{\Phi}\ket{X}
\]
with 
\[
\ket{\Phi} = \sum_{m}A_{m}\ket{\phi_{m}}, \,\, \ket{X} = \sum_{n}B_{n}\ket{\chi_{n}}
\]
so we have, as usual, $C_{\bem} = A_{m}B_{n}, C^{*}_{\bk} = A^{*}_{k}B^{*}_{\ell}$.  Then
\begin{eqnarray*}
\text{Tr}^{b}(\mathcal{P}_{\Psi}) &=& \sum_{mnk}A_{m}B_{n}A^{*}_{k}B^{*}_{n}\ket{\phi_{m}}\bra{\phi_{k}} \\
&=& \sum_{mk}A_{m}A^{*}_{k}\ket{\phi_{m}}\bra{\phi_{k}} = \ket{\Phi}\bra{\Phi} \\
&=& \mathcal{P}_{\Phi}
\end{eqnarray*}
as expected. 

In the entangled case, it suffices to consider the simplest entangled state
\begin{equation}
\label{pse3}
\ket{\Psi_{e}} = \alpha \ket{\Psi^{1}} + \beta \ket{\Psi^{2}}
\end{equation}
where the two states on the right are product states, 
\[
\ket{\Psi^{p}} = \ket{\Phi^{p}}  \ket{X^{p}}, \,\, p = 1,2
\]
The projector is 
\[
\mathcal{P}_{e}\equiv \mathcal{P}_{\Psi_{e}} = |\alpha |^{2}\mathcal{P}_{\Psi^{1}} + |\beta |^{2}\mathcal{P}_{\Psi^{2}} + \alpha \beta^{*} \ket{\Psi^{1}}\bra{\Psi^{2}} +  \alpha^{ *}\beta \ket{\Psi^{2}}\bra{\Psi^{1}} 
\]
To compute the partial trace we need only
\begin{eqnarray*}
\text{Tr}^{b}(\ket{\Psi^{p}}\bra{\Psi^{q}}) & = & \sum_{s}\bra{\chi_{s}}\ket{\Psi^{p}}\bra{\Psi^{q}}\ket{\chi_{s}} \\
&=&\ket{\Phi^{p}}\bra{\Phi^{q}}\bra{X^{q}}(\sum_{s}\ket{\chi_{s}} \bra{\chi_{s}}) \ket{X^{p}} 
\end{eqnarray*}
Since the parenthesized factor is unity, we conclude
\begin{equation}
\label{pqp}
\text{Tr}^{b}(\ket{\Psi^{p}}\bra{\Psi^{q}}) =  \braket{X^{q}| X^{p}} \ket{\Phi^{p}}\bra{\Phi^{q}}
\end{equation}
whence
\begin{equation}
\text{Tr}^{b}(\mathcal{P}_{e}) = |\alpha |^{2}\mathcal{P}_{\Psi^{1}} + |\beta |^{2}\mathcal{P}_{\Psi^{2}} 
+ (\alpha \beta^{*}\braket{X^{2}|X^{1}} \ket{\Phi^{1}}\bra{\Phi^{2}} + h.c.)\label{pse2}
\end{equation}
where $h.c.$ stands for the Hermitian conjugate of the preceding term. 

Notice that \begin{enumerate} \item The partial trace of a pure entangled state in $\mathcal{H}$ is a mixed state in $\mathcal{H}_{a}$.  \item If the entangled states are orthogonal in $\mathcal{H}_{b}$,
\[
\braket{X^{2}|X^{1}} = 0
\]
then the partial trace carries no information from $\mathcal{H}_{b}$. But otherwise the partial trace has a residue of dependence on the $\mathcal{H}_{b}$ states $X^{p}$. In this regard, we recall (subsection \ref{subpro}) that measurement always produces orthogonal states.\end{enumerate}

As an example of a partial trace of a pure entangled state we consider the projector $\mathcal{P}_{\Psi_{e}}$ onto the state $\Psi_{e}$ of (\ref{tan}).  Since $\braket{\chi_{m}|\Psi_{e}} = \delta_{m1}\ket{\phi_{1}}+ \delta_{m2}\ket{\phi_{2}}$ we find the partial trace
 \[
 \sum_{m}\braket{\chi_{m}| \mathcal{P}_{\Psi_{e}}|\chi_{m}} = \ket{\phi_{1}}\bra{\phi_{1}} + \ket{\phi_{2}}\bra{\phi_{2}}
 \]
 to be, as anticipated, the density operator for a mixed state.

\paragraph{Density trace}
Since the density operator, in proper form, is simply a linear combination of projectors, its trace is easily written down using (\ref{pse2}).  Nonetheless is may be useful to display an alternative formalism.

For a product state we know that $\bro = \bro^{a}\bro^{b}$ and it is easy to show that
\begin{equation}
\label{tab}
\text{Tr}^{b}(\bro^{a}\bro^{b}) = \bro^{a}
\end{equation}
More generally we have (\ref{gr}), which we now express as
\begin{equation}
\label{srpp}
\bro = \sum_{\bem' \bem''}\rho_{\bem' \bem''}\ket{\psi_{\bem''}}\bra{\psi_{\bem'}}
\end{equation}
where
\[
\rho_{\bem' \bem''} = \sum_{\bem}P_{\bem}C^{\bem}_{\bem'}C^{*\bem}_{\bem''}
\]
For the trace we use a special case of (\ref{pqp}):
\[
\text{Tr}(\ket{\psi_{\bem''}}\bra{\psi_{\bem'}}) = \delta_{nn'}\ket{\phi_{m''}}\bra{\phi_{m'}}
\]
Thus 
\[
\bro^{a} = \sum_{m'm''}\rho^{a}_{m'm''}\ket{\phi_{m''}}\bra{\phi_{m'}}
\]
with
\begin{equation}
\label{rcc}
\rho^{a}_{m'm''} = \sum_{\bem}P_{\bem}\sum_{n'}C^{\bem}_{m'n'}C^{*\bem}_{m''n'}
\end{equation}
It can be seen that (\ref{tab}) is reproduced in the unentangled case, since then the $n'$-sum contains the factor 
\[
\sum_{n'}B_{n'}B^{*}_{n'} = 1
\]
Otherwise there will occur one or more entangled states, with coefficients  
\[
C_{mn} = \alpha A^{1}_{m}B^{1}_{n} + \beta A^{2}_{m}B^{2}_{n}
\]
Such terms will contribute
\[
\sum_{n'}C_{m'n'}C^{*}_{m''n'} = |\alpha |^{2} A^{1}_{m'}A^{1}_{m''} + |\beta |^{2} A^{2}_{m'}A^{2}_{m''} + (\alpha \beta^{*}A_{m'}A^{2}_{m''}\bar{B}+ h.c.)
\]
where
\begin{equation}
\label{orth2}
\bar{B} \equiv \sum_{n'}B^{1}_{n'}B^{*2}_{n'}  = \braket{X_{2}|X_{1}} 
\end{equation}
and we see again that, unless all entangled states involve only orthogonal states in subsystem $b$, the trace will involve the state of that subsystem.

\section{Evolution of bipartite systems}
\subsection{Density operator evolution}

The bipartite versions of (\ref{qt1}) and (\ref{qt2}) differ from those equations in obvious and simple ways. Thus regarding (\ref{qt1}), we can now express the Hamiltonian for a bipartite system as $H = H_{a} + H_{b} + V_{ab}$, where the last term describes interaction between the two systems. Note that, regardless of any entanglement, the two subsystems are \emph{dynamically} linked.  Each will be modified by interaction with the other. 

With regard to (\ref{qt2}), we consider first a simple example.  Suppose the initial state is al pure state, but an entangled one, given by (\ref{tan}), and that only one of the two systems is ``measured.''   Thus we take the projector $\mathcal{R}_{m}$ to refer to a measurement in system $b$ alone, $\mathcal{R}_{m} = \ket{\chi_{m}}\bra{\chi_{m}}$. Note that the states $\chi_{m}$, belonging to a Hermitian operator, are necessarily orthogonal. We will denote the initial density by $\bro_{0} \equiv \bro(t=0)$
\[
\bro_{0}= \ket{\Psi_{e}} \bra{\Psi_{e}}
\]
Then
\[
\bro(t) = \sum_{m}\ket{\chi_{m}}\bra{\chi_{m}} \braket{\chi_{m}|\Psi_{e}} \braket{\chi_{m}|\Psi_{e}}^{*}
\]
Now
\[
\braket{\chi_{m}|\Psi_{e}}  = \delta_{m1}\ket{\phi_{1}} + \delta_{m2}\ket{\phi_{2}}
\]
so we have
\begin{equation}
\label{ret}
\bro(t) = \mathcal{P}_{\phi_{1}} \mathcal{P}_{\chi_{1}} + \mathcal{P}_{\phi_{2}} \mathcal{P}_{\chi_{2}} = \mathcal{P}_{\phi_{1}\chi_{1}} + \mathcal{P}_{\phi_{2}\chi_{2}} 
\end{equation}
This result could have been anticipated from (\ref{evp});  it differs only in allowing for both possible outcomes of the measurement.  But $\bro(t)$ is very different from $\bro_{0}$; in particular it no longer describes a pure state. 

Next we consider the general case.  The initial, pre-measurement density operator is
\[
\bro_{0}  =  \sum_{\bem \bk}\rho_{0\bem \bk}\ket{\Phi_{0\bem}}\bra{\Phi_{0\bk}}
\]
with $\ket{\Phi_{0\bem}} = \ket{\phi_{0m}}\ket{\chi_{0n}}$, \emph{etc}. Here and below the indices $m$ and $k$ belong to system $a$, and the indices $n$ and $\ell$ belong to system $b$. After measurement we have
\begin{equation}
\label{rpt}
\bro(t) = \sum_{\bem}p_{\bem}\mathcal{R}_{\bem}
\end{equation}
with
\[
\mathcal{R}_{\bem} = \mathcal{R}_{m} \mathcal{R}_{n}  = \ket{\phi_{m}} \ket{\chi_{n}} \bra{\phi_{m}} \bra{\chi_{n}}
\]
Note that the eigenstates here cannot be freely chosen, but are determined by the experimental apparatus.  We substitute the expressions for $\mathcal{R}_{\bem}$ and $\bro_{0}$ into (\ref{qt2}), 
\[
\bm{\rho}(t) = \sum_{\bem}\mathcal{R}_{\bem} \bro_{0} \mathcal{R}_{\bem}
\]
and after straightforward manipulation find that (\ref{rpt}) is produced, with
\begin{equation}
\label{pcc}
p_{\bem} = \sum_{\bem' \bk'}\rho_{0\bem' \bk'}A^{m}_{m'k'}B^{n}_{n'\ell'}
\end{equation}
where
\begin{eqnarray*}
A^{m}_{m'k'} &=& \braket{\phi_{m} | \phi_{0m'}}\braket{\phi_{0k'} | \phi_{m}} \label{amc}\\
B^{n}_{n' \ell'} &=& \braket{\chi_{n} | \chi_{0n'}} \braket{\chi_{0\ell'} | \chi_{n}} \label{bmc}
\end{eqnarray*}
We can interpret these equations as displaying the components of matrices $\bA^{m}$ and $\bB^{n}$; according to our index convention $\bA^{m}$ ($\bB^{n})$ acts only on subsystem $a$ ($b$). With this understood, (\ref{pcc}) can be expressed as a double contraction:
\begin{equation}
\label{pcc2}
p_{\bem}  = \bA^{m} : \bro_{0} : \bB^{n}
\end{equation}
and (\ref{rpt}) becomes
\begin{equation}
\label{punch2}
\bro(t) = \sum_{\bem}\mathcal{R}_{\bem}\bA^{m} : \bro_{0} :  \bB^{n}
\end{equation}

We have noted that the density matrix for a bipartite system cannot in general be expressed as a product, $\bro = \bro^{a}\bro^{b}$, and our analysis nowhere assumes such an expression.  Yet (\ref{punch2}) shows that the operator advancing $\bro$ in time, through measurement, \emph{can} be expressed as such a product.

The condition
\[
\text{Tr}(\bro) = 1 = \sum_{\bem}p_{\bem}
\]
is obviously satisfied by (\ref{rpt}).  It also follows from (\ref{pcc}), since normalization of the eigenkets implies
\begin{equation}
\label{sab}
\sum_{m} A^{m}_{m'k'} = \delta_{m'k'} = \sum_{n}B^{n}_{m'k'}
\end{equation}

\subsection{Partial trace evolution}\label{pte}
Equation (\ref{punch2}) gives the density operator for an ``outside'' observer, who views both subsystems.  The operator appropriate to an observer $A$ who has no knowledge of possible measurements on subsystem $b$ is
\begin{eqnarray*}
\bro^{a}(t) = \text{Tr}^{b}(\bro) &=&  \sum_{\bem}\text{Tr}^{b}(\bA^{m} : \bro_{0} : \bB^{n}\mathcal{R}_{\bem}) \\
&=& \sum_{m}\ket{\phi_{m}}\bra{\phi_{m}}\bA^{m}:\bro_{0}:\sum_{ns}\bB^{n}\bra{\chi_{s}}\ket{\chi_{n}}\bra{\chi_{n}}\ket{\chi_{s}}
\end{eqnarray*}
It may be helpful to write this last expression in terms of indices:
\[
\text{Tr}^{b}(\bro) = \sum_{m}\ket{\phi_{m}}\bra{\phi_{m}}\sum_{m'k'n'\ell'}A^{m}_{m'k'}\rho_{0m'n'k'\ell'}\mathcal{B}_{n'\ell'}
\]
where, in view of (\ref{sab}),
\[
\mathcal{B}_{n'\ell'} \equiv \sum_{ns}B^{n}_{n'\ell'}\delta_{ns}^{2} = \sum_{n}B^{n}_{n'\ell'} = \delta_{n'\ell'}
\]
Hence we have \cite{weinberg2015}
\begin{equation}
\label{pu1}
\text{Tr}^{b}(\bro) = \sum_{m}\ket{\phi_{m}}\bra{\phi_{m}}\sum_{m'k'n'}A^{m}_{m'k'}\rho_{0m'n'k'n'}
\end{equation}

For a shorter version of this result we note that the trace over $b$-indices $(n,\ell)$ of the matrix $\bro_{0}$ is
\[
\text{Tr}^{b}(\rho_{0m'n'k'\ell'}) = \sum_{n'}\rho_{0m'n'k'n'}
\]
Note that this trace removes information about which $b$-system states entered the original density.  Now we can write
\begin{equation}
\label{punch7}
\bro^{a} = \sum_{n} p^{a}_{n} \mathcal{R}_{n} 
\end{equation}
where
\begin{equation}
\label{ra}
p^{a}_{n} =  \bA^{m} : \text{Tr}^{b}(\bro_{0}) 
\end{equation}
Here the orthogonality of the states $\chi_{m}$ has avoided the terms $\bar{B}$ of (\ref{orth2}).

Of course there is an analogous expression for $\bro^{b}$.  Significantly, the partial trace has removed all information about the traced system, regardless of entanglement.  

When two systems are entangled, an observation of subsystem $a$ can immediately change the state of subsystem $b$---a disturbing challenge to relativity (despite the fact that, as we have shown, it cannot be used to transmit any message from $a$ to $b$).  We now see in (\ref{ra}) that no such change occurs in the density operator.  This circumstance suggests that the density operator is the physical structure in quantum theory, the state being merely a mathematical convenience.\cite{weinberg2014}

\section{Generalized linear evolution}
An entirely general, linear evolution law may be expressed by a linear operator, $g$, with \cite{gisin1989}
\[
\mathcal{P}_{\phi_{i}(t)} = g\cdot \mathcal{P}_{\phi_{i}(0)}
\]
for some set of projectors 
\[
\mathcal{P}_{\phi_{i}(t)} = \ket{\phi_{i}(t)}\bra{\phi_{i}(t)}
\]
The proper density operator,
\[
\bro(0) = \sum_{i} p_{i}\mathcal{P}_{\phi_{i}(0)}
\]
then evolves according to
\[
\bro(t) = \sum_{i} p_{i}\mathcal{P}_{\phi_{i}(t)}=  g\cdot \bro_{0}
\]
In general  $\bro$ will have a distinct proper expression:
\[
\bro(0) = \sum_{i} r_{i}\mathcal{P}_{\chi_{i}(0)}
\]
Since the states $\chi_{i}(0)$ may be expressible in terms of the $\phi_{i}(0)$, linearity implies
\[
\mathcal{P}_{\chi_{i}(t)} = g\cdot \mathcal{P}_{\chi_{i}(0)}
\]
whence, again, 
\begin{equation}
\label{rpp2}
\bro(t) =  g\cdot \bro_{0}
\end{equation}
This rule is said to describe a \emph{closed} evolution law for the density operator; we have found that linearity implies, and is clearly implied by, closure.

An evolution law that is not closed might depend upon the representation of $\bro$; that is, we might have
\[
\sum_{i} r_{i}\mathcal{P}_{\chi_{i}(0)} = \sum_{i} p_{i}\mathcal{P}_{\phi_{i}(0)}
\]
while
\[
\sum_{i} r_{i}\mathcal{P}_{\chi_{i}(t)} \neq \sum_{i} p_{i}\mathcal{P}_{\phi_{i}(t)}
\]

\setcounter{chapter}{4}
\chapter{Entropy}\label{Ch4ENT}
\section{Von Neumann entropy}
Von Neumann\cite{vonneumann1955} introduced the entropy 
\begin{equation}
\label{defs0}
S(\rho) = -\text{Tr}(\bro \log \bro)
\end{equation}
Since the density operator is expressed in terms of it's eigenstates $\ket{\phi_{m}}$ by
\[
\bro = \sum_{m}p_{m}\ket{\phi_{m}}\bra{\phi_{m}} = \sum_{m}p_{m}\mathcal{P}_{m}
\]
we have from (\ref{fa}) 
\[
\bro \log \bro = \sum_{m} (p_{m}\log p_{m})\, \mathcal{P}_{m} 
\]
and therefore
\begin{equation}
\label{s2}
S(\rho) = - \sum_{i}p_{i}\log p_{i} = -\braket{\log p_{i}}
\end{equation}
It is clear that this quantity cannot be negative.  It vanishes only for a pure state:
\[
S(\mathcal{P}) = \log 1 = 0
\]
In this sense the von Neumann entropy measures ignorance of the system state.

\section{Comparison to Boltzmann entropy}
Statistical mechanics (SM) in its fully quantum version is based on the distinction between microstates, which are normally taken to be energy eigenstates (stationery states), and macrostates, described by averages over the ensemble.  Thus if $E$ is the exact (micro-)energy of a member of the ensemble, then the thermodynamic system energy (``internal energy'') of the system is
\[
U = \braket{E}
\]
The Boltzmann entropy $\sigma$ is a macrovariable measuring the number $\Omega$ of microstates within a given macrostate:
\[
\sigma  = \log \Omega
\]

The essential assumption of equilibrium statistical mechanics is that, in an isolated system, all microstates consistent with the macrovariables are equally likely $P_{SM} = \text{constant}$.  From this assumption one can determine the probability distribution of a non-isolated system---namely, a system in contact with a reservoir.  The simplest reservoir, called a ``heat bath,'' exchanges energy with the system.  Assuming only that the reservoir is sufficiently large, SM finds that the probability distribution becomes proportional to the Boltzmann factor
\[
P_{SM} \propto \me^{-E/\tau}
\]
where $\tau$ is the temperature in energy units.  More explicitly it is shown that
\[
P_{SM} = \me^{(U - E - \tau \sigma)/\tau}
\]
Since $\braket{U - E} = 0$, this relation implies that
\begin{equation}
\label{slp}
\sigma = -\braket{\log P_{SM}}
\end{equation}
This formula is sometimes called the ``Gibbs entropy.'' It was used repeatedly by Einstein to study fluctuations.

The obvious resemblance between (\ref{slp}) and (\ref{s2}) belies their very different meanings.  In particular, while the $p_{i}$ in (\ref{s2}) are essentially free parameters,  (\ref{slp}) holds only for the probability distribution $P_{SM}$ that has been deduced from statistical physics.  Furthermore, in applications of the von Neumannn entropy, the sum rarely includes all possible system states.  Finally, we note that the probability of a single microstate, $P_{SM}$ must be---unlike that $p_{i}$ in (\ref{s2})---extremely small; indeed
\[
\braket{\log P_{SM}} = -\log \Omega
\] 
and SM requires $\Omega$ to be huge.

The von Neumann entropy might better be compared with the Shannon entropy, which is given by the same formula, with $p_{i}$ now representing the probability of some event (such as the receipt of a message or the toss of a coin).

\section{Additivity of entropy}
Consider a system that is conveniently decomposed into two subsystems, labelled by $a$ and $b$.  We noted in subsection \ref{prodsec} that when no state of system $a$ is entangled with a state in system $b$ then the density operator for the combined system takes the form
\[
\bro = \bro^{a}\bro^{b}
\]
Then we have
\[
S = -\text{Tr}[ \bro^{a}\bro^{b}(\log \bro^{a} + \log \bro^{b})]
\]
It suffices to consider just the first term, using the notation of section \ref{bs}:
\begin{eqnarray*}
\text{Tr} (\bro^{b}\bro^{a}\log \bro^{a})  &= &\sum_{m,n}\bra{\chi_{m}}\bra{\phi_{n}}\bro^{b}\bro^{a} \log \bro^{a}\ket{\phi_{n}}\ket{\chi_{m}}\\
&=&\sum_{m}\braket{\chi_{m}|\bro^{b}|\chi_{m}} \sum_{n}\braket{\phi_{n}|\bro^{a} \log \bro^{a}|\phi_{n}}
\end{eqnarray*}
Here the first factor, involving $\bro^{b}$, is unity because of the unit-trace condition.  Since the second factor is the trace of $\bro^{a}\log \bro^{a}$, we have, in an obvious notation,
\[
S = S^{a} + S^{b}
\]
In this case the entropy is additive.\cite{weinberg2015}

\section{Time evolution}\label{sevol}
We have
\begin{equation}
\label{dsdt}
\frac{dS}{dt} = -\text{Tr}[(1 + \log \bro)\frac{d\bro}{dt}] = -\text{Tr}(\log \bro \,\frac{d\bro}{dt})
\end{equation}
Here we noted that $\text{Tr}(d\bro/dt) = 0$, since $\text{Tr}(\bro) = 1$ must be preserved.  

In the case of an isolated system, we recall (\ref{rt}),
\[
\p_{t} \bm{\rho} = -\mi [H, \bm{\rho}]
\]
in order to write
\[
\frac{dS}{dt} =  \mi \text{Tr}(\log \bro [H, \bm{\rho}])
\]
This quantity vanishes, as we would expect, since the trace of a product of three Hermitian operators is invariant under arbitrary rearrangements.  (Recall subsection \ref{proalg}.)

Entropy evolution of a non-isolated system is studied in Chapter \ref{Ch6LIN}.

\setcounter{chapter}{5}
\chapter{The Bell Inequalities}\label{Ch5BEL}

\section{Correlated fermions} \label{sec:label}
\subsection{Generalized Pauli states}

Since this chapter will use formulae regarding two-state (spin) systems repeatedly, we begin by reviewing some algebra of such systems.

Let 
\[
\bn = \hat{x}n_{x} + \hat{y}n_{y} + \hat{z}n_{z} 
\]
be a unit vector and
\[
n_{\perp} \equiv n_{x} + \mi n_{y}
\]
Then the operator for measurement of fermionic spin in the direction $\bn$ is
\begin{equation}
\label{defsn}
\sigma_{n} \equiv \bn \cdot \bm{\sigma} = \left(\begin{array}{cc} n_{z} & n_{\perp}^{*} \\ n_{\perp} & -n_{z} \end{array}\right)
\end{equation}
It can be seen that $\sigma_{n}$ reduces to the familiar Pauli operators,
\begin{eqnarray}
\sigma_{x} &=& \left(
\begin{array}{cc}
  0 & 1   \\
  1 & 0\\
\end{array}
\right),  \\
\sigma_{y} &=& \left(
\begin{array}{cc}
  0 & -\mi   \\
  \mi & 0\\
\end{array}
\right) \\
\sigma_{z} &=& \left(
\begin{array}{cc}
  1 & 0   \\
  0 & -1\\
\end{array}
\right) 
\end{eqnarray}
in the appropriate limits.  It is also easily confirmed that the eigenvalues of $\sigma_{n}$ are $\pm1$.  Denoting the eigenvectors by $\ket{s_{n} \pm}$, a straightforward calculation yields
\begin{eqnarray}
\ket{s_{n}+} &=& \sqrt{\frac{1 + n_{z}}{2}}\left(\begin{array}{c} 1 \\ \frac{n_{\perp}}{1+ n_{z}}\end{array}\right) = 2^{-1/2}\left(\begin{array}{c} \sqrt{1 + n_{z}} \\ \frac{n_{\perp}}{\sqrt{1+ n_{z}}}\end{array}\right)  \\
\ket{s_{n }-} &=& \sqrt{\frac{1 - n_{z}}{2}}\left(\begin{array}{c} 1 \\ \frac{-n_{\perp}}{1- n_{z}}\end{array}\right) = 2^{-1/2}\left(\begin{array}{c} \sqrt{1 - n_{z}} \\ \frac{-n_{\perp}}{\sqrt{1- n_{z}}}\end{array}\right)
\end{eqnarray}
Thus $\ket{s_{n} \pm}$ represents a state in which the spin, measured in the direction $\bn$, has value $\pm 1$.

These kets must reduce to the familiar kets
\begin{eqnarray}
\ket{x+} &=& 2^{-1/2}
\left(
\begin{array}{c}
  1    \\
  1 \\
\end{array}
\right), \,\,\, \ket{x-} = 2^{-1/2}
\left(
\begin{array}{c}
  1    \\
  -1 \\
\end{array}
\right)\label{x-} \\
\ket{y+} &=& 2^{-1/2}
\left(
\begin{array}{c}
  1    \\
  \mi \\
\end{array}
\right), \,\,\, \ket{y-} = 2^{-1/2}
\left(
\begin{array}{c}
  1    \\
  -\mi \\
\end{array}
\right) \\
\ket{z+} &=& 
\left(
\begin{array}{c}
  1    \\
  0 \\
\end{array}
\right), \,\,\, \ket{z-} = 
\left(
\begin{array}{c}
  0    \\
  1 \\
\end{array}
\right)
\end{eqnarray}
in appropriate limits.  Because of the singularity at $n_{z} = \pm1$, the reduction requires care.  Thus we consider the component
\[
X_{-} \equiv \frac{-n_{\perp}}{\sqrt{2(1 - n_{z})}}
\]
In spherical polar coordinates
\[
X_{-} = \frac{-\sin \theta(\cos \phi + \mi \sin \phi)}{\sqrt{2(1- \cos \theta)}}
\]
Now suppose $\phi = 0$ and $\theta =  \ep$, and let $\ep \rightarrow 0$. Then $\cos \theta \rightarrow 1 - \ep^{2}/2 $ and $\sin \theta \rightarrow \ep$, so we find
\[
X_{-} \rightarrow -1
\]
in agreement with $\ket{x_{-}}$ in (\ref{x-}).

\subsection{Entanglement in Hilbert space}
We consider the standard situation: two fermions are correlated in a singlet state, resulting from the decay of a spin-$0$ source.  The fermions have become widely separated spatially, unable to interact.  Only the spin values of the two fermions matter, so the system space is four-dimensional.  As a basis we use the eigenstates of the spin's $z$-components, with kets $\ket{+} = \ket{z+}$ and $\ket{-} = \ket{z-}$. The singlet state of the bipartite system is given by
\begin{equation}
\label{defp}
\ket{\psi} = 2^{-1/2}(\ket{+}_{1}\ket{-}_{2} - \ket{-}_{1}\ket{+}_{2})
\end{equation}
This state is unique up to phase.

The spin operators for the two fermions are denoted by $\bm{\sigma}_{1}$ and $\bm{\sigma}_{2}$.   The boldface refers to the vector character of each $\bm{\sigma}$, which can have an arbitrary orientation in three-dimensional coordinate space $(x,y,z)$.   Introducing the unit vectors $\ba$ and $\bb$, we consider the operator product of $\bm{\sigma}_{1}\cdot \ba$ and $ \bm{\sigma}_{2}\cdot \bb$:
\[
S \equiv (\bm{\sigma}_{1}\cdot \ba) (\bm{\sigma}_{2}\cdot \bb)
\]
which is a scalar in coordinate space, but four-dimensional in the Hilbert space.

It can be convenient to use row and column vectors.  Thus the first term of (\ref{defp}) is
\[
\ket{+}_{1}\ket{-}_{2}  = \left(\begin{array}{c} 1 \\ 0 \end{array}\right) \\ \otimes \left(\begin{array}{c} 0 \\ 1 \end{array}\right)= \left(\begin{array}{c} 0 \\1\\0\\0 \end{array}\right)
\]
the second term is
\[
\ket{-}_{1}\ket{+}_{2} = \left(\begin{array}{c}0\\0\\1\\0 \end{array}\right)
\]
so we have
\begin{equation}
\label{pc}
\ket{\psi} = 2^{-1/2}\left( \begin{array}{c} 0 \\1\\-1\\0 \end{array}\right)
\end{equation}

\subsection{Eigenvalues and eigenvectors}
We choose our $z$-axis along the unit vector $\bb$, so that $\bm{\sigma}_{2} \cdot {\bb} = \sigma_{2z}$ and $S = (\bm{\sigma}_{1} \cdot  \ba) \sigma_{2z}$. We then recall (\ref{defsn}):
\[
\bm{\sigma}_{1} \cdot \ba = \left(
\begin{array}{cc} a_{z} & a_{\perp}^{*} \\  a_{\perp} & -a_{z}
\end{array} \right)
\]
where $a_{\perp} = a_{x} + \mi a_{y}$. Thus
\begin{equation} \label{defsm}
S = \left(
\begin{array}{cccc}
a_{z} & 0 & a_{\perp}^{*} &0 \\
0& -a_{z} &0 & - a_{\perp}^{*} \\
a_{\perp}  &0&-a_{z}&0 \\
0& -a_{\perp} &0 &a_{z}
\end{array}\right)
\end{equation}

The eigenvalues are denoted by $\lambda$ and found from
\[
\det(S - \lambda \bI) = 0
\]
Noting that $a_{z}^{2} + |a_{\perp}|^{2} = 1$, we obtain the eigenvalue equation
\[
(\lambda^{2} - 1)^{2} = 0
\]
Hence there are two eigenvalues, 
\[
\lambda = \pm 1
\]
each doubly degenerate. The normalized eigenvectors are, for $\lambda = -1$,
\begin{eqnarray}
V_{1}&=& \sqrt{\frac{1 - a_{z}}{2}}\left(0, \frac{a_{z}+1}{a_{\perp}},0 , 1\right) \\
V_{2} &=&  \sqrt{\frac{1 + a_{z}}{2}}\left(\frac{a_{z} - 1}{a_{\perp}}, 0,1,0 \right)
\end{eqnarray}
and, for $\lambda = 1$,
\begin{eqnarray}
V_{3}&=&  \sqrt{\frac{1 + a_{z}}{2}}\left(0, \frac{a_{z}-1}{a_{\perp}},0 , 1\right) \\
V_{4} &=&  \sqrt{\frac{1 - a_{z}}{2}}\left(\frac{a_{z} + 1}{a_{\perp}}, 0,1,0 \right)
\end{eqnarray}

\subsection{Statistics of singlet state}
\paragraph{Expectation value}
The expectation value of $S$ in the state $\ket{\psi}$ is
\[
\braket{S} = \braket{\psi | S |\psi}
\]
Expressing $\ket{\psi}$ using (\ref{pc}) we easily find
\begin{equation}
\label{sp}
S \cdot \ket{\psi} = 2^{-1/2}\left( \begin{array}{c} -a^{*}_{\perp} \\-a_{z}\\a_{z}\\-a_{\perp} \end{array}\right)
\end{equation}
whence
\begin{equation}
\label{punch1}
\braket{S} =  \frac{1}{2} \left( \begin{array}{cccc} 0 & 1& -1& 0 \end{array}\right)\left( \begin{array}{c} -a_{\perp}^{*} \\-a_{z}\\a_{z}\\-a_{\perp} \end{array}\right) = -a_{z} = -\ba \cdot \bb
\end{equation}

\paragraph{Variance (quantum uncertainty)}
We denote variance of $S$ by $\Delta_{S}$:
\[
\Delta_{S}= \braket{(S - \braket{S})^{2}}
\]
We compute it from the eigenvectors $V_{n}$, using
\[
(S - \braket{S})^{2}\ket{V_{n}}    = (\lambda_{n} - \braket{S})^{2} \ket{V_{n}} 
\]
and 
\[
\ket{\psi} = \sum_{n}\braket{V_{n}| \psi}\ket{V_{n}}
\]
Thus 
\begin{equation}
\label{ds2}
\Delta_{S} = \sum_{n} |\braket{V_{n}| \psi}|^{2}(\lambda + a_{z})^{2}
\end{equation}

Easy calculations give, for $n = 1,2 \,\,(\lambda = -1)$,
\[
|\braket{V_{n}| \psi}|^{2} = \frac{1 + a_{z}}{4} 
\]
and for $n = 3,4 \,\,(\lambda = 1)$,
\[
|\braket{V_{n}| \psi}|^{2} = \frac{1 - a_{z}}{4} 
\]
Thus (\ref{ds2}) becomes
\[
\Delta_{S} = \frac{1}{2} [(1 + a_{z})(1 - a_{z})^{2} + (1 - a_{z})(1 + a_{z})^{2}] = 1-a_{z}^{2}
\]
There is no uncertainty when $\ba \cdot \bb = \pm1$ and maximum uncertainty when $\ba \cdot\bb = 0$.

\paragraph{Probability distribution}
Consider the bipartite state $\phi$ in which $\bm{\sigma}_{1}$ yields positive spin in the $z$-direction, correspoding to the ket $\ket{z+}_{1}$, and measurement of $\bm{\sigma}_{2}$ yields positive spin in the $\bb$-direction, which deviates by the angle $\al$ from the vertical.  We have found that the ket for subsystem $2$ is  
\[
\ket{s_{b}+} = \ 2^{-1/2}\left(\begin{array}{c} \sqrt{1 + b_{z}} \\ \frac{b_{\perp}}{\sqrt{1+ b_{z}}}\end{array}\right) 
\]
Thus the full system state is $\ket{\phi} = \ket{z+}_{1}\ket{s_{b}+}_{2}$.  If the initial bipartite state is the singlet $\ket{\psi}$, then the state $\ket{\phi}$ will be observed with probability $P(\al) = |\braket{\psi | \phi}|^{2}$, where
\[
\braket{\psi | \phi} =  \frac{1}{\sqrt{2}}\bra{z+}_{1}\bra{s_{b}+}_{2}(\ket{z+}_{1}\ket{z-}_{2} - \ket{z-}_{1}\ket{z+}_{2}) = \frac{1}{2}\frac{b_{\perp}}{\sqrt{1 + b_{z}}}
\]
Thus we have
\begin{eqnarray}
P(\al) &=& \frac{|b_{\perp}|^{2}}{4(1+ b_{z})} \\
&=& \frac{1}{4} \frac{\sin^{2} \al}{(1 + \cos \al)} \\
&= & \frac{1}{2}\sin^{2}(\frac{\al}{2}) \label{ps2}
\end{eqnarray}

This is a monotonic function, vanishing when $\al = 0$ (because the spins must be opposite) and reaching its maximum of $1/2$ (because the spins can have any orientation) at $\al = \pi/2$.  

\subsection{Density operators}
The singlet state is pure, so its density operator is the projector
\[
\mathcal{P}^{\psi} = \ket{\psi}\bra{\psi}
\]
The density operator in the two-dimensional system of fermion $1$, appropriate to an observer who is ignorant of subsystem $2$, is the partial trace
\[
\bro_{1} = \text{Tr}^{2}(\mathcal{P}^{\psi}) = \bra{+}_{2}\ket{\psi} \bra{\psi}\ket{+}_{2} +\bra{-}_{2}\ket{\psi} \bra{\psi}\ket{-}_{2}
\]
and similarly for $\bro_{2}$.  Now (\ref{defp}) gives
\[
\bra{\pm}^{2}\ket{\psi} = \frac{\mp 1}{\sqrt{2}}\ket{\mp}_{1}
\]
whence
\begin{equation}
\label{dab}
\bro^{1} =  \frac{1}{2}(\mathcal{P}^{+}_{1} + \mathcal{P}^{-}_{1})
\end{equation}
where  $\mathcal{P}^{\pm} = \ket{\pm}\bra{\pm}$, and similarly for $\bro_{2}$. We notice that \begin{enumerate} \item As usual, the partial trace of a pure, entangled state is a mixed state.  \item The partial traces for each system contain no information about the other system.\end{enumerate}

\section{Alice and Bob}
\subsection{Classic setup}
Alice and Bob, separated by many light years, each measure the spin direction of incoming fermions. The spin operators are denoted as before by $\bm{\sigma}_{1,2}$.  We denote the orientation of Alice's (Bob's) detector by the unit vector $\ba$ ($\bb$). A decaying particle with zero spin emits a pair of fermions, one to Alice and one to Bob, in a singlet state.  Thus both Alice and Bob measure, with their respective orientations, the state $\psi$ of (\ref{defp}). A final element in the set-up is a tabulation, made by some third observer, of the two orientations and the two observations for each emitted fermion. In the \emph{delayed choice} scenario, at least one observer chooses an orientation $\ba$ or $\bb$ after the fermions are launched. Good general references on the following discussion are \cite{BellBook} and \cite{mermin1993}.

\subsection{Correlations}
Clearly if Alice measures the $z$-component of her fermion and finds $+1$, then Bob is certain to find his fermion has $z$-component $-1$.  This correlation would pertain classically and is entirely unremarkable. It results from the common origin of the two fermions---that is, from the fact that they \emph{did} interact in the past.  One says that correlation can result from ``common cause.'' 

Quantum mechanics yields a much more interesting correlation: Bob's observations are correlated with the orientation of Alice's apparatus.  According to (\ref{punch1}),  
\begin{equation}
\label{qmc}
C(\ba, \bb) \equiv \langle (\bm{\sigma}_{1} \cdot \ba)  (\bm{\sigma}_{2} \cdot \bb) \rangle = -\ba \cdot \bb
\end{equation}
This correlation appears regardless of when, during the flight of the particles, the orientations are chosen. Since the two observers are space-like separated, the correlation must be viewed as shocking. It is hard to believe that the orientations chosen by Alice and Bob can have a common cause.

\subsection{Special relativity}
Since the separation between Alice and Bob is space-like, the temporal order of their observations depends upon the observer: the correlation cannot be related to cause and effect.  In fact the correlation challenges the time-space structure associated with special relativity.  Some information about the orientations, however they may have been chosen or changed, is available at both observer sites.

Yet the nature of this common information does not allow superluminal communication between Alice and Bob.  The correlation is detectable only to an observer who is aware of both orientations and both eigenvalues.  For example, if Alice chooses the orientation $\ba = \hat{z}$ and observes $+1$, then the state of Bob's fermion is $\ket{+}\ket{-}$, so if Bob happens to set $\ba = \hat{z}$ he will certainly observe $-1$. But \emph{he won't know that the result was certain}.  The impossibility of communication is also evident in the Alice's and Bob's density operators, as found in subsection \ref{pte}. 

There is an apparent communication trick: suppose that $\ba$ remains fixed while many fermions are sent to Bob.  One might think that the statistics of these fermions would be revealing, but they are not.  The point is that while all the fermions correspond to the same setting, they do not correspond to the same observation.  Alice observes, in the long run, half spin-up and half spin-down, for any setting, so when Bob counts the cases when $\bb$ happens (without his knowledge) to equal $-\ba$, he will see the same $50\%$ statistics and find nothing special about his setting. 

Bob would see revealing statistics only if he could make many copies---``clones''---of the incoming state, before performing his measurement.  Each clone would then correspond to not only the same orientation $\ba$ but also the the same observation by Alice.  Bob could then perform experiments with various orientations $\bb$ on this collection and deduce the orientation $\ba$.  For example Bob would find that a certain orientations $\bb$ ( = $\pm \ba$) always produced the same result.  Thus cloning would permit superluminal communication.  However Bob can perform, according to quantum mechanics, only linear operations on a state, and we show next that cloning is not a linear operation.\cite{noclone}

\subsection{No-cloning theorem}
To clone a state $\psi$ from a Hilbert space $\mathcal{H}$, one needs a copy of $\mathcal{H}$ to store the clone, so we must work in the Hilbert space $\mathcal{H}\bigotimes\mathcal{H}$.  The cloning operator  $C$ is then defined by
\begin{equation}
\label{cu}
C (\ket{\psi}\ket{u}) = \me^{\mi \phi}\ket{\psi}\ket{\psi}
\end{equation}
where $u$ is an arbitrary state and $\phi$ is a phase.  But quantum mechanical operations correspond to linear operators, and no linear operator $C$ could satisfy (\ref{cu}).  Indeed, if $\psi = A + B$ then
(\ref{cu}) implies
\[
C\ket{\psi}\ket{u} =  \me^{\mi \phi}(\ket{A}+\ket{B})(\ket{A} +\ket{B}) 
\]
while linearity requires
\[
C\ket{\psi}\ket{u} = C\ket{A}\ket{u} + C\ket{B}\ket{u}  =  \me^{\mi \phi_{A}}\ket{A}\ket{A} + \me^{\mi \phi_{B}}\ket{B}\ket{B} 
\]
Since the second form lacks terms $\ket{B}\ket{A}$ and $\ket{A}\ket{B}$ there is no linear operator that clones.

\subsection{Not really}
A measurement by Alice or Bob, neither of whom knows what the other may have measured, cannot be predicted by quantum mechanics; the quantum state can give at most statistical predictions.  A plausible supposition is that the indeterminacy results, as in classical statistical mechanics, from lack of information: there exists information, not contained in the quantum state yet carried by each fermion, that determines what will be observed for each orientation $\ba$ and $\bb$.  The point of Bell's inequality is that this supposition, associated with \emph{hidden variables} or \emph{realism}, is inconsistent with the predictions of quantum mechanics.  The quantum correlation formula (\ref{qmc}) does not allow any pre-measurement assignment of values for the fermion spins. Indeterminacy is much deeper than that resulting from lack of knowledge.\cite{mermin1993}

The situation is analogous to the well-known indeterminacy in the double-slit experiment, where the supposition that each particle went through one slit or the other can be shown to be quantum mechanically untenable.

To show that pre-experimental values are not only unknown but undetermined we suppose that Alice and Bob use only two orientations, denoted by $(\ba_{0}, \ba_{1})$ and $(\bb_{0}, \bb_{1})$.  Then consider the operator, in the full bipartite Hilbert space,
\[
X \equiv (A_{0} + A_{1})B_{0} + (A_{0} - A_{1})B_{1}
\]
Here we use the abbreviations $A_{i} \equiv \ba_{i} \cdot \bm{\sigma_{1}}$, $B_{i} \equiv \ba_{i} \cdot \bm{\sigma_{2}}$, with $i = (0,1)$. 

We now assume that, before any measurement, each spin component has a definite, if unknown, value---the assumption of realism. From momentum conservation and the quantum mechanical fact that the only possible values for each spin are $\pm 1$, we see that there are just two possibilities: \begin{enumerate} \item $A_{0} + A_{1} = 2, \,\,A_{0} - A_{1} = 0$ \item $A_{0} -A_{1} = 2, \,\,A_{0} +A_{1} = 0$  \end{enumerate} Since the value of $X$ is both cases is either zero or two, realism requires the expectation value of $X$ to satisfy 
\[
\braket{X} \leq 2
\]
Next we consider the choice
\[
\ba_{0} = \hat{z}, \,\, \ba_{1} = \hat{x}, \,\, \bb_{0} = -(1/\sqrt{2})(\hat{x} + \hat{z}), \,\, \bb_{1} = (1/\sqrt{2})(\hat{x} - \hat{z})
\]
and compute the quantum mechanical value of $\braket{X}$.  From (\ref{qmc}) we find
\[
\braket{A_{0}B_{0}} = 1/\sqrt{2}, \,\, \braket{A_{0}B_{1}} = 1/\sqrt{2}, \,\, \braket{A_{1}B_{0}} = 1/\sqrt{2}, \braket{A_{1}B_{1}} = -1/\sqrt{2}
\]
and therefore 
\[
\braket{X}  = 2\sqrt{2}
\]
Thus realism cannot pertain.

There are many ways to illustrate the failure of realism---the need for a  ``contextual''  description.  In particular, we can view the Stern-Gerlach device as a filter, by blocking any particle with (say) downward orientation: only spin-up particles pass through.  Consider the angular span $0 < \al < \pi/2$ as decomposed into two equal segments $0 < \al < \pi/4$ and $\pi/4 < \al < \pi/2$.  The probability for passage through the full span is at most $P(\pi/2)$; the \emph{maximum} probability for passage through each of the two segments is $P(\al/4)$ (these two probabilities are the same by rotational symmetry).  Since, in a realistic view, any particle passing the full span must passed one of the two segments, we conclude
\[
P(\pi/2) \leq 2P(\pi/4)
\]
But this realist picture is contradicted by the quantum mechanical formula (\ref{ps2}): $P(\pi/2) = 0.25 > 2P(\pi/4) = 0.146$.   (A qualitatively identical argument pertains for any division of the full span into segments.)
 
 An additional demonstration, ``GHZ-Mermin,'' involves three detectors, which, along with the source, lie in the $x-y$ plane. \cite{greenberger1989,mermin1994} The three detectors are labelled $(\ba,\bb,\bc)$, and each much be aligned along either the $x$ or $y$ axis, so the data for any experiment will consist of three numbers $\{a_{x,y}, b_{x,y},c_{x,y} \}$.  Each number can be only $\pm1$.  Realism, or non-contexuality, assumes that these numbers are ``elements of reality``---that they exist, although unknown, prior to any measurement.  In that case the following manipulations make sense.
 
Suppose that the source prepares the following tri-partite state (recall $\ket{\pm} \equiv \ket{z\pm}$):
 \[
 \ket{\psi} = 2^{-1/2}(\ket{-} \ket{-} \ket{-}  - \ket{+} \ket{+} \ket{+})
 \]
 This is normalized:
 \[
 \braket{\psi| \psi} = \frac{1}{2}(\bra{-}\bra{-}\bra{-} \ket{-} \ket{-} \ket{-}  +\bra{+}\bra{+}\bra{+} \ket{+} \ket{+} \ket{+} ) = 1
 \]
We use the relations
\begin{equation}
\label{xyz}
\sigma_{x}\ket{\pm} = \ket{\mp},\,\,\, \sigma_{y}\ket{\pm} = \mi \ket{\mp} 
\end{equation} 
to show that $\ket{\psi}$ is an eigenket of $S_{xyy}  \equiv \sigma_{x}\sigma_{y}\sigma_{y}$, with eigenvalue $+1$.
That is
\[
S_{xyy} \ket{-} \ket{-} \ket{-} = - \ket{+} \ket{+} \ket{+}
\]
and
\[
S_{xyy} \ket{+} \ket{+} \ket{+} = - \ket{-} \ket{-} \ket{-}
\]
whence
\[
S_{xyy}\ket{\psi} = \ket{\psi}
\]
The same argument shows that $S_{yxy}$ and $S_{yyx}$ are also eigenstates with eigenvalue $+1$.  Since the eigenvalue of $S_{ijk}$ is $a_{i}b_{j}c_{k}$, we have derived the identities
\[
a_{x}b_{y}c_{y}  = a_{y}b_{x}c_{y} = a_{y}b_{y}c_{x} = 1
\]
Next consider the product
\[
(a_{x}b_{y}c_{y})(a_{y}b_{x}c_{y})(a_{y}b_{y}c_{x}) = 1 = a_{x}b_{x}c_{x}a_{y}^{2}b_{y}^{2}c_{y}^{2} 
\]
Since each squared factor is unity, we conclude
\begin{equation}
\label{real}
a_{x}b_{x}c_{x}  = 1
\end{equation}
This directly contradicts the quantum mechanical result
\begin{equation}
\label{qms}
S_{xxx}\ket{\psi} = -\ket{\psi}
\end{equation}
which follows immediately from (\ref{xyz}).  

The point again is that the three numbers $a,b,c$ cannot be assigned values until the observations are made; the observations brings these numbers into existence.  A special feature of the GHZ-Mermin argument is that the only probabilities that occur are $0$ and $1$.  The QM result (\ref{qms}) and the realistic result (\ref{real}) are both certain.

\setcounter{chapter}{6}
\chapter{The Linblad equation}\label{Ch6LIN}

\section{Introduction}
Standard quantum mechanics allows for two types of evolution:  \begin{enumerate} \item Hamiltonian (Schr\"odinger) evolution of an isolated system. \item The ``jump'' of a system subject to measurement by an external agent.  \end{enumerate}  The evolution law in both cases is linear and closed; that is, in both cases
\begin{equation}
\label{lc0}
\bro(t + \Delta t) = \bM \cdot \bro(t)
\end{equation}
for some operator $\bM$. (In the Schr\"odinger case $\bM$ is unitary.)

It can be argued that neither evolution law applies in the common case of a system subject to constant interaction with a noisy environment.  Mainly for this reason, we seek a third quantum evolution law.  To avoid various complications (including the ability to send superluminal messages) we continue to demand linearity and closure, beginning with (\ref{lc0}). We then seek the most general form of the operator $\bM$ that preserves \begin{enumerate} \item Hermiticity, $\bro^{\dagger} = \bro$; \item The trace condition $\text{Tr}(\bro) = 1$; \item Positivity $\braket{v |\bro | v} \geq 0$ for any state $\ket{v}$. (We will find that this last condition requires slight modification.) \end{enumerate} The general linear operator that satisfies these conditions was first found by Lindblad \cite{linblad1976} and independently by Gorini et al.\cite{sudarshan1976}.  See also \cite{banks1984}.   It is sometimes called the GKSL equation, but more commonly the Lindblad equation.  Our derivation follows the very clear treatment by Pearle \cite{pearl2012}.  

The Lindblad equation has three applications: \begin{enumerate} \item The description of the jump process, linking an isolated system to a measured system. \item The description of a system interacting with a noisy environment.  \item As a proposed replacement for the Schr\"odinger equation.  \end{enumerate}  The first of these applications bears of course on the question of density operator collapse.  In fact Weinberg\cite{weinberg2016} has shown that the collapse seen in (\ref{qt2}) emerges naturally, as a smooth evolution, from the Lindblad equation: no jump. That is, returning to the comments at the end of subsection \ref{qevol}, we can say that the first phase of density collapse becomes continuous under Lindblad dynamics. 

After a brief comment on algebra, we find in Section \ref{two}  the form of $\bM$ that satisfies Hermi\-ticity and unit trace.  The positivity constraint, which is more complicated, is resolved in Section \ref{three}.

\section{Algebra review} 
\subsection{Hermitian matrices}
Any matrix can be expressed as a sum of dyads; for a Hermitian matrix $\bM$ this sum has a peculiar form:
\[
\bM = \sum_{k}\alpha_{k}\bA^{k}\bA^{k \dagger} 
\]
where the $\alpha_{k}$ are real numbers and the $\bA^{k}$ are linearly independent vectors.  Being independent, the $A^{k}$ can be linearly combined and rescaled to form an orthonormal set.  Denoting the orthonormal vectors by $\bE^{k}$, we have
\begin{equation}
\label{mee}
\bM  = \sum_{k} \lambda_{k}\bE^{k}\bE^{k\dagger}
\end{equation}
Now observe that the $\bE^{k}$ are simply the eigenvectors, with eigenvalues $\lambda_{k}$, of $\bM$:
\[
\bM \cdot \bE^{k} = \sum_{k} \lambda_{k}\bE^{k}\delta_{kk} = \lambda_{k}\bE^{k}
\]
(Here and below we denote the ``inner'' matrix product with a dot.) It may happen that the linear space has dimension $N$ while the $\bE^{k}$ span only a subspace of dimension $M < N$.  In that case $\bM$ is Hermitian in the larger space but has $N-M$ zero eigenvalues.

We conclude that a Hermitian matrix can always be expressed in terms of its eigenvalues $\lambda_{k}$ and normalized eigenvectors $\bE^{k}$ as the sum (\ref{mee}).  In component form
\begin{equation}
\label{mee2}
M_{mn} = \sum_{k} \lambda_{k}E^{k}_{m}E^{k*}_{n}
\end{equation}

\subsection{Matrices as vectors}
The space of square matrices with non-vanishing determinants is closed under algebraic operations and thus forms a vector space. If the standard vector space is denoted by $S$, then a square matrix $\bA$ can be viewed as a vector in the product space $S \bigotimes S$.  In fact the square-matrix vector space differs from the usual vector space in only one important way: the former is closed under multiplication.  For any tensors $\bA$ and $\bB$ of order $N$ we can construct the tensor product $\bA\bB$, of order $2N$, and then contract to produce a tensor of order $2N - 2$.  But the case $N=2$ is exceptional: only this case does contraction produce a matrix of order $N$.   (Thus the product of two square matrices is another square matrix.)  This multiplicative feature has no effect on the standard vector properties.

We define the inner product of two such matrices $\bA$ and $\bB$ to be the contracted product $\bA \colon \bB$:
\[
 \bA \colon \bB = \sum_{mn}A_{mn}B_{nm} = \text{Tr}(\bA \cdot \bB) 
\]
Linear transformation of the matrix $\bA' = \bM \cdot \bA$ sums over both matrix indices:
\begin{equation}
\label{apa}
A'_{mn} = \sum_{k\ell}M_{mkn\ell}A_{k\ell} 
\end{equation}
To define the Hermitian adjoint of $\bM$ we suppose $\bA$ and  $\bA'$ are both Hermitian.  Then
\begin{eqnarray}
A'^{*}_{nm} &=& \sum_{k\ell}M^{*}_{nkm\ell} A^{*}_{k\ell} \nonumber \\
&=& \sum_{k\ell}M^{*}_{nkm\ell} A_{\ell k}  \nonumber\\
&=& \sum_{k\ell}M^{*}_{n \ell mk} A_{k \ell}  = A'_{mn} \label{apa2}
\end{eqnarray}
Comparing (\ref{apa}) and (\ref{apa2}) see that $\bM$ is Hermitian iff
\begin{equation}
\label{defher}
M_{mkn\ell} = M_{n\ell mk}^{*}
\end{equation}
Using two-component vector subscripts, $\bn = n \ell,\,\, \bem = m k$, we can write the Hermiticity condition as
\begin{equation}
\label{mh}
M^{*}_{\bn \bem} = M_{\bem \bn}
\end{equation}

In an alternative approach, matrix vectors can be mapped onto ordinary vectors by ordering the indices, according to, for example, 
\[
11, 12, \cdots, 21, 22, \cdots
\]
Thus a square $N\times N$ matrix becomes a vector with $N^{2}$ components, and all of the usual vector theory applies.\cite{pearl2012} Of course this mapping obscures the multiplicative feature.

\section{Linear transformation of $\bro$} \label{two}
\subsection{Preserving hermiticity}
Here we suppose that two density operators, $\bro$ and $\bro'$, are linearly related:  $\bro' = \bM \cdot \bro$. Since the components of $\bM$ form a four-component matrix, we need to extend the formalism of the previous section.  A natural definition for vector multiplication provides the relation between components,
\begin{equation}
\label{rmr}
\rho'_{mn}  = \sum_{k\ell} M_{mkn\ell}\rho_{k\ell}
\end{equation}
Since $\bro$ is Hermitian, (\ref{rmr}) implies
\[
\rho'^{*}_{nm} = \sum_{k\ell} M^{*}_{nkm\ell}\rho^{*}_{k \ell} = \sum_{k\ell} M^{*}_{nkm\ell}\rho_{\ell k}
\]
Requiring that $\bro'$ also be Hermitian, and interchanging the dummy indices $k$ and $\ell$, yields
\[
\sum_{k\ell} (M^{*}_{n\ell mk} - M_{mkn\ell})\rho_{k\ell} = 0
\]
Thus, in view of (\ref{defher}), we see that Hermiticity of the density is preserved whenever the transformation matrix $\bM$ is Hermitian.

Hermiticity implies that $\bM$ has eigenmatrices,  satisfying $\bM \cdot \bE^{i} = \lambda_{i} \bE^{i}$; explicitly
\[
\sum_{k\ell}M_{mkn\ell}E_{k\ell}^{i} = \lambda_{k}E^{i}_{mn}
\]
where the $\lambda_{k}$ are real and the matrices $\bE^{k}$ are orthogonal.  Then the $\bE^{k}$ can be normalized to produce the orthonormality condition
\begin{equation}
\label{eed}
\bE^{k}\colon \bE^{\dagger \ell} = \text{Tr}(\bE^{k}\cdot \bE^{\dagger \ell}) =  \delta_{k\ell}
\end{equation}

We next extend the argument leading to (\ref{mee}) to the eigenmatrix case.  The components of the fourth-rank Hermitian matrix $\bM$ can be expressed in terms of its eigenmatrices as
\begin{equation}
\label{mee4}
M_{mkn\ell} = \sum_{i} \lambda_{i} E^{i}_{mk} E^{i*}_{n \ell}
\end{equation}
or
\[
M_{\bem \bn} = \sum_{i} \lambda_{i} E^{i}_{\bem} E^{i*}_{\bn}
\]
Here the Hermiticity, $M_{\bem \bn}  = M^{*}_{\bn \bem}$,   is manifest.  
Note that there are $N^{2}$ eigenvalues and eigenmatrices; our notation is abbreviated:
\[
\sum_{k} \equiv \sum_{k}^{N^{2}}
\]

Returning to (\ref{rmr}), we now have
\begin{equation}
\label{rmr2}
\rho'_{mn} = \sum_{ik\ell} \lambda_{i} E^{i}_{mk} E^{i*}_{n \ell}\rho_{k\ell}
\end{equation}
This relation has a straightforward matrix representation:
\begin{equation}
\label{punch10}
\bro' = \sum_{i} \lambda_{i} \bE^{i}\cdot \bro \cdot \bE^{i\dagger}
\end{equation}
Here the dots represent ordinary matrix multiplication in the original space of dimension $N$. 

We conclude that, if the matrices $\bro$ and $\bro'$ are \begin{enumerate} \item linearly related, and \item both Hermitian \end{enumerate} then their relation can be expressed by (\ref{punch10}), in which the $\bE^{i}$ form an arbitrary set of orthonormal matrices.  
%
\subsection{Trace condition}
Next we require that $\bro'$, like $\bro$, have unit trace.  We have
\begin{eqnarray}
\text{Tr}(\bro') &= &\sum_{k}\lambda_{k}\text{Tr}(\bE^{k}\cdot \bro \cdot \bE^{k\dagger}) \nonumber \\
&=& \sum_{k}\lambda_{k}\text{Tr}(\bE^{k\dagger} \cdot\bE^{k}\cdot \bro) \label{ltee}
\end{eqnarray}
We introduce the Hermitian operator
\[
\bT \equiv \bI - \sum_{k}\lambda_{k}\bE^{k\dagger} \cdot\bE^{k} 
\]
Here $\bI$ is the unit operator, again in the original space of dimension $N$. Then, since $\text{Tr}(\bro') = \text{Tr}(\bro)$, we can express (\ref{ltee}) as
\begin{equation}
\label{tr0}
\text{Tr}(\bT \cdot \bro) = 0
\end{equation}
Notice that (\ref{trq0}) states that the expectation value of $\bT$ vanishes for \emph{any} density operator, 
\[
\braket{\bT} = 0
\]
and recall that this implies $\bT = 0$.  That is
\begin{equation}
\label{punch20}
\sum_{i}\lambda_{i}\bE^{i\dagger} \cdot\bE^{i} = \bI
\end{equation}
More explicitly
\begin{equation}
\label{punch30}
\sum_{i}\sum_{n}\lambda_{i} E^{i\dagger}_{mn}  E^{i}_{nk} = \delta_{mk}
\end{equation}

The trace of (\ref{punch20}) is of interest. Recall from (\ref{eed}) that the trace $\text{Tr}(\bE^{i\dagger} \cdot\bE^{i}) = 1$, while $\text{Tr}(\bI) = N$, the dimension of the space.  Therefore
\begin{equation}
\label{tree}
\sum_{i}\lambda_{i} = N
\end{equation}

\section{Positivity}  \label{three}
\subsection{Complete positivity}
We need to show that $\braket{v|\bro' |v} \geq 0$ for every state $\ket{v}$.  We use (\ref{punch10}) to write
\[
\braket{v|\bro' |v} = \sum_{k}\lambda_{k}\braket{v |\bE^{k} \cdot \bro \cdot \bE^{k\dagger} |v}
\]
and then define the state
\[
\ket{u_{k}} \equiv \bE^{k\dagger}\ket{v}
\]
Then
\[
\braket{v|\bro' |v} = \sum_{k}\lambda_{k} \braket{u^{k}| \bro |u^{k}}
\]
which, since $\bro$ is a positive operator, is clearly non-negative if all the eigenvalues $\lambda_{k}$ are non-negative. 

Significantly, the converse statement is false: since the weights $ \braket{u^{k}| \bro |u^{k}}$ can be freely chosen, a positive operator could have one or more negative eigenvalues.  We will find that the Lindblad derivation requires positive eigenvalues, so at this point we require
\begin{equation}
\label{lpi}
\lambda_{k} \geq 0
\end{equation}
for all $i$.  The mathematical requirement is that the density operator is not only positive, but also \emph{completely positive}.  Complete positivity can be shown to be a weak additional requirement on $\bro$.
\paragraph{Krauss representation}

Complete positivity allows us to define $\bK^{k} \equiv \sqrt{\lambda_{k}}\bE^{k}$ and thus to write
\begin{equation}
\label{kra}
\bro' = \sum_{k}\bK^{k} \cdot \bro \cdot \bK^{k\dagger}
\end{equation}
with the requirement
\begin{equation}
\label{kra2}
\sum_{k}\bK^{k} \cdot \bK^{k\dagger} = \bI
\end{equation}

\paragraph{Special case}
In an important special case, the transformation matrix $\bM$ is the identity, $\bro' = \bro$.  From (\ref{mee4}) we see that in this case
\[
 \sum_{i} \lambda_{i} E^{i}_{mk} E^{i*}_{n \ell} = \delta_{mk}\delta_{\ell n}
 \]
Multiplying both sides by $E^{j}_{\ell n}$, summing over $\ell$, and using the orthogonality condition (\ref{eed}) we obtain
\[
 \lambda_{j}E^{j}_{mk}  = E^{j}_{nn} \delta_{km} = \text{Tr}(\bE^{j}) \delta_{km}
\]
That is, any eigenmatrix with non-vanishing eigenvalue must be proportional to a unit matrix.  Since the orthogonal set of eigenmatrices can include only one unit matrix, all the other eigenvalues must vanish. Recalling (\ref{tree}), we conclude that the identity $\bro = \bro'$ case is characterized by a single non-vanishing eigenvalue, say $\lambda_{1} = N$, whose associated eigenmatrix is a multiple of the unit matrix: $\bE^{1} = a\bI$. Then (\ref{punch10}) becomes $\bro' = \bro  = Na^{2}\bro$, so $a = N^{-1/2}$ and
\[
\bE^{1} = N^{-1/2}\bI
\]
 in the case of identity transformation.  We emphasize that the $\bE^{k}$ for $k \neq 1$ are unspecified; they form an arbitrary set of orthonormal matrices, constrained only by (\ref{punch20}).

\subsection{Summary}
We have found that the general closed, linear transformation of the density operator can be expressed as (\ref{punch10}), 
\begin{equation}
\label{puch1}
\bro' = \sum_{k} \lambda_{k} \bE^{k}\cdot \bro \cdot \bE^{k\dagger}
\end{equation}
where the real numbers $\lambda_{k}$ and matrices $\bE^{k}$ are constrained \emph{only} by the equations
\begin{eqnarray}
\lambda_{k} &\geq& 0, \text{all} \, i \label{l+}\\
\sum_{k}\lambda_{k} &=& N \label{N+}\\
\text{Tr}(\bE^{k}\cdot \bE^{\dagger \ell}) &= & \delta_{k\ell} \\
\sum_{k}\lambda_{k}\bE^{k\dagger} \cdot \bE^{k} &=& \bI \label{lee}
\end{eqnarray}

\section{Lindblad evolution}
\subsection{Perturbation theory}\label{linblad}
The Lindblad equation prescribes evolution of an open system under four assumptions:  \begin{enumerate} 
\item Closure: evolution of the density operator depends only on that operator, without reference to quantum states.  \item  Markov: the density operator $\bro(t + \delta t)$ depends only on $\bro(t)$. \item Linearity: the relation between $\bro(t + \delta t)$ and $\bro(t)$ is linear. \item Differentiability: the derivative 
\[
\frac{d\bro	}{dt}  = \bro(t + \delta t) - \bro(t) + \mathcal{O}(\delta t^{2})
\]
exists. \end{enumerate}
It follows from our discussion that these assumptions imply
\begin{equation}
\label{lin0}
\bro(t + \delta t) = \sum_{k} \lambda_{k}(t+ \delta t) \bE^{k}(t+ \delta t)\cdot \bro(t) \cdot \bE^{k\dagger}(t+ \delta t)
\end{equation} 
Now observe that
\begin{eqnarray}
\lambda_{k}(t + \delta t) &=& \delta_{k1}N +\dot{\lambda}_{k}\delta t \\
\bE^{1}(t + \delta t) &=& N^{-1/2}\bI +\dot{ \bE^{1}}\delta t \\
\bE^{k}(t + \delta t) &=& \bE^{k}(t) + \mathcal{O}(\delta t), \,\,\, k \neq 1
\end{eqnarray}
where the dots indicate time derivatives. Substitution into (\ref{lin0}) yields
\begin{equation}
\label{lin2}
\frac{d\bro}{dt} = N^{1/2}(\dot{\bE}^{1}\cdot \bro + \bro \cdot \dot{\bE}^{1 \dagger} ) + \sum_{k}\dot{\lambda}_{k}\bE^{k} \cdot \bro \cdot \bE^{k\dagger}
\end{equation}
where all quantities on the right-hand side are evaluated at time $t$. Next we evaluate (\ref{N+}) at $t + \delta t$ to find 
\[
\dot{\lambda}_{1}\delta t = -\sum_{k\neq 1}\lambda_{k}(t + \delta t)
\]
It then follows from (\ref{l+}) that $\dot{\lambda}_{1} \leq 0$.  At this point we simplify notation by writing
\begin{eqnarray}
\dot{\lambda}_{1} &=& -cN \\
N^{1/2}\dot{\bE}^{1} &=& \bB
\end{eqnarray}
Thus (\ref{lin2}) becomes
\begin{equation}
\label{lin3}
\frac{d\bro}{dt} =  - c\bro +\bB\cdot \bro + \bro \cdot \bB^{\dagger}  + \sum_{k\neq 1}\dot{\lambda}_{k}\bE^{k} \cdot \bro \cdot \bE^{k\dagger}
\end{equation}

Consider next (\ref{lee}); in the present notation
\[
(\bI - c \delta t)[ \bI + \delta t(\bB^{\dagger} + \bB)] + \sum_{k \neq 1}\lambda_{k}(t+\delta t)\bE^{k\dagger} \cdot \bE^{k} = \bI
\]
The first-order terms are
\[
-c \bI + \bB^{\dagger} + \bB + \sum_{k \neq 1}\dot{\lambda}_{k}\bE^{k\dagger} \cdot \bE^{k} = 0
\]
We pre- and post-multiply this relation by $\bro$ and sum the results to find
\begin{equation}
\label{cr2}
-c \bro + \frac{1}{2}(\bro \cdot \bB^{\dagger} + \bB^{\dagger}\cdot\bro + \bro\cdot\bB + \bB \cdot\bro) + \frac{1}{2}\sum_{k \neq 1}\dot{\lambda}_{k}(\bro\cdot\mathcal{E}_{k} + \mathcal{E}_{k} \cdot\bro) = 0
\end{equation}
where we abbreviate
\[
\mathcal{E}_{k} = \bE^{k\dagger} \cdot \bE^{k} 
\]
Subtraction of (\ref{cr2}) from (\ref{lin3}) yields
\begin{equation}
\label{lin5}
\frac{d \bro}{dt} = \frac{1}{2}[(\bB - \bB^{\dagger}), \bro] + \sum_{k}\dot{\lambda}_{k}[\bE_{k}\cdot \bro \cdot \bE_{k}^{\dagger} - \frac{1}{2}(\bro \cdot \mathcal{E}_{k}  + \mathcal{E}_{k} \cdot \bro)]
\end{equation}
where the first term on the right is the commutator defined by
\[
[\bA, \bB] \equiv \bA \cdot \bB - \bB \cdot \bA = \bA\bB - \bB\bA
\]
Here and below we suppress the dot in operator products.  

The result (\ref{lin5}) is essentially the Lindblad equation.  However it is conventional to modify the notation.  First observe that, for any operator $\bB$, $\bB - \bB^{\dagger}$ is an anti-Hermitian operator.  We therefore define the Hermitian operator $\bH \equiv (\mi/2)(\bB - \bB^{\dagger})$, which can play the role of a Hamiltonian.  Then, in terms of the operators $\bL^{k} \equiv \sqrt{\dot{\lambda}_{k}} \bE_{k}$, (\ref{lin5}) becomes 
\[
\frac{d \bro}{dt} = \mathcal{L} \bro
\] 
where $\mathcal{L}$ is the Lindblad operator
\begin{equation}
\label{lin6}
\mathcal{L} \bro = -\mi [\bH, \bro] +\sum_{k\neq 1}[\bL^{k} \bro   \bL^{k\dagger} - \frac{1}{2}(\bL^{k}\bL^{k\dagger}\bro+ \bro \bL^{k}\bL^{k\dagger})] 
\end{equation}
Notice that trace preservation
\[
\text{Tr}(d\bro) = \text{Tr}(\mathcal{L} \bro) dt = 0
\]
is manifestly satisfied: commutators always have vanishing trace, and all the combinations of $\bro L^{k}L^{k}$ have equal traces, so that they add to zero.  (Recall the formulae of subsection \ref{proalg}.) Hermiticity of the product $\mathcal{L} \bro$ is also evident.
   
\subsection{Lindblad solutions}
We denote the eigen-matrix solutions of the linear operator $\mathcal{L}$ by $q^{\alpha}$ and the eigenvalues by $\lambda^{\al}$:
\[
\mathcal{L} q^{\alpha} = \lambda^{\alpha}q^{\alpha}
\]
Since as noted above, the left-hand side of this equation has zero trace for any matrix $q^{\al}$, we see that
\begin{equation}
\label{trq0}
\lambda^{\al} \neq 0 \Rightarrow \text{Tr}(q^{\al}) = 0
\end{equation}
So the $q^{\al}$ are not in general density operators.  However it is easy to construct, from the $q^{\al}$, the density operator $\bro$ that solves (\ref{lin6}).  Introducing\[
Q^{\al}\equiv \me^{\lambda^{\al}t}q^{\al}
\]
we see that
\[
\frac{d Q^{\al}}{d t} = \lambda^{\al}\me^{\lambda^{\al}t}q^{\al} = \mathcal{L}Q^{\al}
\]
is an exact Lindblad solution for each $\al$, whence\cite{weinberg2015} 
\begin{equation}
\label{rsq}
\bro = \sum_{\al}Q^{\al}
\end{equation}
is a solution---in fact the general solution when the eigen-matrices $q^{\al}$ are complete.  In view of (\ref{trq0}), we expect most of the terms in (\ref{rsq}) to be traceless.  But, to allow $\text{Tr}(\bro)= 1$, there must be at least one vanishing eigenvalue.

\section{Entropy change}
\subsection{General form}
A physical evolution operator should not allow the (von Neumann) entropy,
\[
S(\bro) = -\text{Tr}(\bro \log \bro)
\]
to decrease. We recall (\ref{dsdt}) to write
\[
\frac{dS}{dt} = -\text{Tr}(\log \bro\,\mathcal{L}\bro)
\]
The first, Hamiltonian, term in (\ref{lin6}) does not contribute here, as shown in section \ref{sevol}. There remains (after mild rearrangement)
\begin{equation}
\label{stll}
\frac{dS}{dt} = - \text{Tr}[\log \bro\,\sum_{k\neq 1}(\bL^{k} \bro \bL^{k\dagger} - \bL^{k}\bL^{k\dagger}\bro)]
\end{equation}
Denote the bracketed quantity by $\bK$; it has components
\[
\bK_{mn} = \sum_{p}(\log \bro)_{mp}\sum_{k\neq 1}(\bL^{k} \bro \bL^{k\dagger} - \bL^{k}\bL^{k\dagger}\bro)_{pn}
\]
We use the eigenvectors of $\bro$ as basis vectors, so that 
\[
\rho_{mn} = \delta_{mn}p_{n}
\]
and, recalling (\ref{fan}),
\[
(\log \bro)_{mp} = \delta_{mp}p_{m}
\]
Thus 
\[
K_{mn} = \log p_{m}\sum_{k\neq 1}(\bL^{k} \bro \bL^{k\dagger} - \bL^{k}\bL^{k\dagger}\bro)_{mn}
\]
We therefore consider the product
\begin{eqnarray*}
(\bL^{k} \bro \bL^{k\dagger})_{mn} &=& \sum_{pq}L^{k}_{mp}\rho_{pq}L^{\dagger k}_{qn} \\
 &= &\sum_{q} p_{q} L^{k}_{mq}L^{k\dagger}_{qn}
\end{eqnarray*}
Similarly 
\[
(\bL^{k}\bL^{k\dagger}\bro)_{mn} = \sum_{q} p_{m} \, L^{k}_{mq}L^{k\dagger}_{qn}
\]
Hence we have
\begin{eqnarray}
\frac{dS}{dt} &=& -\sum_{m}K_{mm} \nonumber \\
&=&- \sum_{mn} \,(p_{n} - p_{m})\, \Lambda_{mn}\log p_{m} \label{stll2}
\end{eqnarray}
where
\[
\Lambda_{mn} \equiv \sum_{k \neq 1}L_{nm}^{k}L_{mn}^{k\dagger} = \sum_{k \neq 1}|L_{mn}^{k}|^{2}
\]
We cannot proceed beyond (\ref{stll}) without additional constraint on the essentially arbitrary operators $\bL^{k}$.

\subsection{Special case}
The case in which the quantities $\Lambda_{mn}$ form a Hermitian matrix is of special physical interest.  After all, the Lindblad equation is supposed to represent a system in contact with some environment, such as an experimental device, so it is natural to expect even the $\bL^{k}$ to be Hermitian.  
Since the $\Lambda_{mn}$ are real, Hermiticity implies symmetry,
\[
\Lambda_{mn} = \Lambda_{nm}
\]
and we can write
\[
\frac{dS}{dt} = - \sum_{mn} \,(p_{m} - p_{n})\, \Lambda_{mn} \log p_{n}
\]
Subtracting this form from (\ref{stll}) we obtain\cite{weinberg2015}\cite{banks1984}
\begin{equation}
\label{punchst}
\frac{dS}{dt}  = \frac{1}{2}\sum_{mn}\, \Lambda_{mn}(p_{n} - p_{m})(\log p_{n} - \log p_{m})
\end{equation}
Then,  since all the components $\Lambda_{mn}$ are positive and $\log x$ is an increasing function of $x$, we have confirmed  
\[
\frac{dS}{dt} \geq 0
\]
Since each term of (\ref{punchst}) is positive, $dS/dt$ can vanish only when all the $p_{n}$ are identical,
\[
p_{n} = 1/N
\]
giving the maximum entropy state $S = \log N$. Aside from this exceptional case, Lindblad evolution causes entropy to increase, in sharp distinction to the Hamiltonian case, discussed in Chapter \ref{Ch4ENT}.

\section{Confronting collapse}
\subsection{Evolution of density operator}
Weinberg \cite{weinberg2016} has shown that the Lindblad equation predicts smooth evolution of the density operator from an arbitrary initial state $\bm{\rho}(0)$ to the state
\[
\bm{\rho}(t) = \sum_{m}\mathcal{R}_{m} \bm{\rho}(0)\mathcal{R}_{m}
\]
Thus the two discordant evolution laws for the density operator, described in Chapter \ref{Ch2DEO}, are replaced by a single dynamical equation.  There is no arbitrary insertion of a new dynamic.  It follows that, if the density operator is viewed as the sole dynamical object of quantum theory, then the Lindblad equation fixes the problem of collapse.  This fix, according to which the density operator is a real physical object while the state vector is at most a computational tool, not requiring dynamical description, is discussed further in subsection \ref{minfix}.  First, however, we mention a different perspective, in which the state vector presumed physical.  Clearly this view requires a  new state-vector dynamic.

\subsection{Spontaneous collapse}
If the state vector is deemed a physical entity, then it should satisfy a well-defined dynamical law---a law that is lacking in conventional quantum mechanics.  Spontaneous collapse theories\cite{GRW1986} provide a law, in the form of a stochastic differential equation that implies intermittent \emph{spatial} collapse of the wave function.  The limitation to collapse in coordinate space is justified by declaring that all physics experiments end with the observation of some spatial location, such as the position of an instrumental pointer. 

The stochastic differential equation forces a single-particle spatial wave function to collapse to a Gaussian of width $\lambda$ at random intervals of average spacing $\tau$.  Because of entanglement, the collapse of any particle's wave function will cause collapse of the entire system, with the result that---while the collapse of isolated micro-systems will be delayed---macro-systems will almost never be found in an uncollapsed state. 

In the present context, the key observation is that spontaneous collapse of the wave function has been shown to imply Lindblad evolution of the density operator: one can view the Lindblad equation as a consequence of spontaneous collapse\cite{GRW1990}.  We will consider a different perspective presently, but here some comments are in order. \begin{enumerate} \item The restriction of collapse to coordinate space---the ``preferred basis''---may be justifiable, but it's fit into standard quantum arguments can be awkward. Can we no longer view, for example, the Stern-Gerlach experiment in terms of a collapse in spin-space? \item The two collapse parameters $\lambda$ and $\tau$ can apparently be chosen consistently with observation, but no physical explanation or derivation of either is provided. \end{enumerate}

\subsection{Theory truncation}

Quantum mechanics resembles a manual for winning at some card game---a book describing the probabilities for certain card combinations.  Of course when the cards are dealt one does not expect the book to explain why a particular hand occurred.  Like that book, quantum mechanics has done its \emph{complete} job when it reveals---based on knowledge of a system's environment and its initial statistics---the likelihood for various possible outcomes: $\bro(0) \rightarrow \bro(t)$.   In subsection \ref{qevol} we found it useful to decompose the experimental process into two phases:  an interaction phase, when some instrument interacts with the system to produce $\bro(t)$, and an observation phase, when the particular outcome is noted.  \emph{Quantum theory pertains exclusively to the interaction phase}. The theory has no more to say about the observed outcome than a card-playing manual has to say about which cards one is dealt.

We have also noted that after the observation a new density operator, now concentrated on the measured data, pertains.  This operator can be used as the $\bro(0)$ in order to follow future evolution of the system.  The quantum role for externally prescribed initial data differs in no significant way from its role in classical physics.

\subsection{The minimal fix}\label{minfix}

The observation that quantum predictions are intrinsically statistical---that the theory cannot predict actual outcomes---of course not new. A novel and plausible response to this limitation is made explicit by Weinberg\cite{weinberg2014}, who suggests that quantum theory should prescribe the evolution of the density operator alone, without concern for the dynamics of state vectors. A similar viewpoint is espoused by Barandes and Kagan\cite{barandes2019}, who label it as ``minimal modal''  and who also grant physical status to a special class of quantum states---the operator's eigenstates. (Clearly an evolution law for $\bro$ implies one for its eigenvectors.)

The description of measurement provided by the Lindblad equation, and its prediction of the Born rule, displays a simple solution to the problem of density operator collapse.  If one assumes that quantum states are mere calculational tools—that only the density operator is physical—then quantum theory loses its troubling flaw. This perspective seems indeed to be the least complicated way to address collapse: the minimal fix to quantum theory.  It can be summarized as follows.
\begin{quote}The central physical object of quantum theory is the density operator.  The state vector plays an important mathematical role, in particular populating the Hilbert space on which the density operator acts, but it is not accorded status as a physical object.  The density operator evolves according to the Lindblad equation---a single law that pertains to an isolated system as well as a measured system.  The dynamics of the state vector is not specified, and not needed to advance the singular goal of quantum theory:  statistical prediction, manifested in the trace formula.  That is, as implicitly understood from the earliest theoretical constructions, quantum mechanics predicts probabilities, not events.  The lack of a well-defiined evolution law for quantum states reflects this predictive limitation.\end{quote}

Several comments are in order. \begin{enumerate}
\item The proper representation of the density operator in terms of non-orthogonal projectors is not unique. This circumstance confirms the non-physical quality of state vectors. The trace formula always provides unique statistical predictions, independent of representation.
\item  In view of the essential role of the Lindblad equation, it is significant that its form is almost devoid of physics (beyond the appearance of the Hamiltonian, as inherited from Schr\"odinger).  The Lindblad equation is derived from little more than the requirement that the density operator, as it evolves, must remain a density operator.
\item By strictly limiting quantum theory to the evolution and predictions of the density operator, one avoids not only the collapse issue but also a host of theoretical complications and controversies, including the role of consciousness, the many-worlds interpretation, and spontaneous collapse theories.  (This simplification is emphasized in the statistical interpretation of Ballentine\cite{Ballentine1970}.)
\item The minimal fix removes, not only any question of collapsing state vectors, but also the conflict between entanglement and relativistic space-time.  We have noted that, in the case of entangled states, a change in the orientation of a measuring device has an immediate effect on a remote state vector---an effect that cannot be used for communication but is nonetheless troubling.  However we have also shown that such a change has no effect on the remote density operator. Since only the density operator is a physical object, any challenge to relativity is eliminated.  
\end{enumerate}

The minimal fix is a scientific surrender; it declares certain aspects of reality to be intrinsically unknowable. It follows Wittgenstein's dictum\cite{witt} ``Whereof one cannot speak, thereof one must be silent.''

\backmatter  
\bibliographystyle{unsrt}
\bibliography{PQW_NEW.bib}

\end{document}